\documentclass[12pt]{article}
\pdfoutput=1 
\usepackage{amssymb, amsthm,amsmath,amsfonts}
\usepackage{graphicx}
\usepackage{xcolor, colortbl}
\usepackage{empheq}
\usepackage[pdftex, bookmarks=true,colorlinks,linkcolor=red,urlcolor=blue, citecolor=blue]{hyperref}

\usepackage{amsthm}
\usepackage{array}
\usepackage{soul}
\usepackage{float}
\usepackage{bm}

\makeatletter\@addtoreset{equation}{section}\makeatother

\newcommand{\preprint}[1]{\begin{table}[t]  
             \begin{flushright}               
             {#1}                             
             \end{flushright}                 
             \end{table}}                     
\renewcommand{\title}[1]{\vbox{\center\LARGE{#1}}\vspace{5mm}}
\renewcommand{\author}[1]{\vbox{\center#1}\vspace{5mm}}
\newcommand{\address}[1]{\vbox{\center\em#1}}

\def\be{\begin{eqnarray}}
\def\ee{\end{eqnarray}}
\def\bea{\begin{eqnarray}}
\def\eea{\end{eqnarray}}
\newcommand{\nn}{\nonumber}

\def\Dslash{\,\,{\raise.15ex\hbox{/}\mkern-12mu D}}
\def\Dbarslash{\,\,{\raise.15ex\hbox{/}\mkern-12mu {\bar D}}}
\def\delslash{\,\,{\raise.15ex\hbox{/}\mkern-9mu \partial}}
\def\delbarslash{\,\,{\raise.15ex\hbox{/}\mkern-9mu {\bar\partial}}}
\def\pslash{\,\,{\raise.15ex\hbox{/}\mkern-9mu p}}
\def\calDslash{\,\,{\raise.15ex\hbox{/}\mkern-12mu {\cal D}}}

\newcommand{\ZZ}{{\mathbb Z}}

\newcommand{\e}{\,{\rm e}}

\newcommand{\bra}{\langle}
\newcommand{\ket}{\rangle}

\def\lae{\mathrel{\mathop{\smash{\lower .5 ex \hbox{$\stackrel<\sim$}}}}}
\def\lae{\mathrel{\mathop{\smash{\lower .5 ex \hbox{$\stackrel>\sim$}}}}}

\def\arXiv#1{\href{http://arxiv.org/abs/#1}{arXiv:#1}}
\def\arXiv#1#2{\href{http://arxiv.org/abs/#1}{arXiv:#1}}

\def\doi#1#2{\href{http://doi.org/#1}{#2}}

\begin{document}

\unitlength = .8mm

\begin{titlepage}
\vspace{.5cm}
\preprint{}

\begin{center}
\hfill \\
\hfill \\
\vskip 1cm

\title{\boldmath 
Quantum Chaos in Topologically Massive Gravity}
\vskip 0.5cm
{Yan Liu}\footnote{Email: {\tt yanliu@buaa.edu.cn}}
and
{Avinash Raju}\footnote{Email: {\tt avinashraju777@gmail.com}}
\address{Center for Gravitational Physics, Department of Space Science, \\ and International Research Institute
of Multidisciplinary Science,
\\ Beihang University,  Beijing 100191, China}
\address{Key Laboratory of Space Environment Monitoring and Information Processing, \\
	Ministry of Industry and Information Technology, Beijing 100191, China}
\end{center}
\vskip 1.5cm

\abstract{
We study quantum chaos of rotating BTZ black holes in Topologically Massive gravity (TMG). We discuss the relationship between chaos parameters including Lyapunov exponents and butterfly velocities from  shock wave calculations of out-of-time-order correlators (OTOC) and from pole-skipping analysis. We find a partial match between pole-skipping and the OTOC results in the high temperature 
regime.  
We also find that the velocity bound puts a chaos  
constraint on the gravitational Chern-Simons coupling. 
}

\vfill

\end{titlepage}

\begingroup 
\hypersetup{linkcolor=black}
\tableofcontents
\endgroup

\section{Introduction} 
\label{sec:intro}

Chaos is ubiquitous to a wide range of physical systems with large number of degrees of freedom and is crucial to thermalization and relaxation of the system. 
An important probe 
of dynamical chaotic behavior 
in quantum many-body systems is 
the exponential growth of certain 
real time out-of-time ordered correlators (OTOC) 
\cite{Shenker:2013pqa, Roberts:2014isa,Shenker:2014cwa,Maldacena:2015waa,Liu:2020rrn}
\begin{eqnarray}
\label{eq:exp}
\frac{\langle V(0)W(t,\vec{x})V(0)W(t,\vec{x}) \rangle_\beta}{\bra V(0)V(0)\ket_\beta \bra W(t,\vec{x})W(t,\vec{x})\ket_\beta} \simeq 1-\varepsilon_{VW}\; \text{exp}\left[\lambda_L\left(t-\frac{|\vec{x}|}{v_B}\right)\right]
\end{eqnarray}
where $V$ and $W$ are generic few-body Hermitian operators, $\beta=1/T$ is the inverse temperature, 
$\lambda_L$ is the quantum Lyapunov exponent, $v_B$ is the butterfly velocity, and $\varepsilon_{VW}$ is a small parameter that typically depends on the  parameters of the theory such as the characteristic energy scale of the operators $V$ and $W$, and the number of degrees of freedom. The exponential growth of \eqref{eq:exp} can be thought of as an operator generalization of classical chaos which measures the sensitivity of a system to the initial conditions. 

One intriguing feature of quantum chaos is the upper bound proposed on the quantum Lyapunov exponent in \cite{Maldacena:2015waa} 
\be \label{eq:bound}
\lambda_L \leq 2\pi/\beta\,,
\ee 
based on very general assumptions of the quantum many-body system such as analyticity and factorization at large time of thermal correlation functions. This upper bound was found to be saturated by black holes and generic large-$N$ theories admitting a black hole dual.  
Moreover, a finite number of higher derivative correction terms of Einstein gravity will not correct the Lyapunov exponent and the chaos bound \eqref{eq:bound} remains true \cite{Maldacena:2015waa, Liu:2020rrn}.  

However the chaos bound (\ref{eq:bound}) found in \cite{Maldacena:2015waa} was shown to be probably  violated in a three dimensional rotating BTZ black hole in Einstein gravity \cite{Stikonas:2018ane,Poojary:2018esz, Jahnke:2019gxr}. As pointed out in \cite{Stikonas:2018ane,Poojary:2018esz,Jahnke:2019gxr} in rotating BTZ black holes there seems to be two Lyapunov exponents, one of which lies strictly within the bound while the other violating the bound
$\lambda_\pm=\frac{2\pi}{\beta (1\mp\Omega\ell)}$ 
where $\Omega$ is the angular velocity of rotating BTZ and $\ell$ is the AdS$_3$ radius. 
However, as emphasised in \cite{Mezei:2019dfv}, after carefully taking into account the periodicity of phase factor, the two coefficients in front of the exponential factors in OTOC are related and there is only one instantaneous Lyapunov exponent. The OTOC shows a modulation on top of the MSS bound and the average Lyapunov exponent remains within the bound. In the high temperature limit, the instantaneous Lyapunov exponent behaves as a step function which jumps between $\lambda_\pm$.\footnote{The chaos bound has also been examined in the presence of a finite chemical potential associated to internal as well as spacetime symmetries \cite{Halder:2019ric}.}

There is also a surprising connection between early time exponential growth of OTOC and late time hydrodynamics which was pointed out first in the numerical studies by \cite{Grozdanov:2017ajz}. 
It was noticed that the retarded two-point function of energy density operator in the complexified momentum space has special pole-skipping points 
\begin{eqnarray}
\label{eq:ps}
\omega_*=i\lambda_L\,, ~~~\quad k_*=\frac{i\lambda_L}{v_B}\,,
\end{eqnarray}
where the correlator is non-unique due to the vanishing residue of the poles at these points. 
More precisely, if we write the correlator as $G^R (\omega,k) = \frac{B(\omega,k)}{A(\omega,k)}$, 
we have $A(\omega_*,k_*) = B(\omega_*,k_*) = 0$ and therefore 
the correlator now depends on the slope $\delta \omega /\delta k$ as we approach $(\omega_*,k_*)$. 
Though OTOCs can be computed explicitly in many specific models, a 
unified, model independent 
understanding of quantum chaos came from this connection and led to the construction of effective theory of maximal   chaos in \cite{Blake:2017ris}.\footnote{The effective theory for chaotic CFTs has been studied in \cite{Haehl:2018izb, Haehl:2019eae}.}  This effective description predicts the existence of pole-skipping  due to the microscopic shift symmetry of the effective hydrodynamic mode. In holography, the special points \eqref{eq:ps} shows up in the bulk because at these 
points the $vv$-component of the near-horizon equations of motion in the ingoing Eddington-Finkelstein coordinates becomes degenerate \cite{Blake:2018leo}.\footnote{Even though the special point was calculated for the energy-density $T^{00}$ correlators, the pole-skipping phenomenon seems to exist for a wide-range of operators 
\cite{Grozdanov:2019uhi,Blake:2019otz,Ceplak:2019ymw}. Other aspects of pole-skipping has been studied in {\it e.g.} \cite{Grozdanov:2018kkt, Natsuume:2019sfp, Li:2019bgc, Ahn:2019rnq,  Das:2019tga, Wu:2019esr, Abbasi:2019rhy}.}
\vskip 0.3cm

The exponential growth of OTOC and pole-skipping for the retarded Green's function of energy density are two different signatures for dynamical quantum chaos. It should therefore be important to understand better the connection between these two different approaches in as wide range of systems as possible. For a class of maximally chaotic systems, these two methods give identical chaotic parameters \cite{Grozdanov:2017ajz, Blake:2017ris, Blake:2018leo}. It is natural to ask, for systems with more than one Lyapunov exponent or non-maximally chaotic systems what are the connections between these two approaches? This motivates us to explore several aspects of quantum chaos for rotating BTZ black holes in 3D gravity using these two different approaches.  

We will first examine the phenomenon of pole-skipping for rotating BTZ in Einstein's gravity. In this specific system the Lyapunov exponents $\lambda_{\pm}$ from OTOC calculations \cite{Stikonas:2018ane,Poojary:2018esz, Jahnke:2019gxr, Mezei:2019dfv} turn out to be robust under pole-skipping, which will be verified from the holographic stress tensor correlator as well as a direct CFT computation of the retarded correlator $G^R_{T^{00}T^{00}}(\omega,k)$. The results from pole-skipping match well with the calculations in \cite{Stikonas:2018ane,Poojary:2018esz, Jahnke:2019gxr} and \cite{Mezei:2019dfv} in the high temperature limit. 

We will also explore quantum chaos for rotating BTZ in Topologically Massive Gravity (TMG) \cite{Deser:1981wh, Deser:1982vy}, \textit{i.e.} three dimensional Einstein gravity deformed by a gravitational Chern-Simons term, using several different approaches. 
Unlike  Einstein's gravity, the linearized fluctuations in TMG has an additional massive, propagating mode. 
Demanding that Lyapunov exponents should satisfy the chaos bound and the butterfly velocity should be less than the speed of light, we find nontrivial constraints on the gravitational Chern-Simons coupling, which is consistent with the black hole stability studies (see {\it e.g.} \cite{Park:2006gt, Li:2008dq} for discussions). In the high temperature limit of rotating BTZ, we show that there is one instantaneous Lyapunov exponent defined using OTOC, which behaves as step functions taking values between two out of the three exponents $\lambda_\pm, \lambda_m$. 
We also study the chaotic parameters using the pole-skipping method, from near horizon equations of motion, and from retarded Green’s function of energy density correlators, as well as from CFT techniques. We find all these pole-skipping methods produce all three independent Lyapunov exponents $\lambda_\pm, \lambda_m$. 

The organization of our paper is as follows. In section \ref{sec:btzreview}, we provide a brief review of rotating BTZ black holes and collect various known facts. In section \ref{sec:eingravity} we study chaos from pole-skipping for a rotating BTZ black hole in Einstein's gravity. 
In section \ref{sec:tmg}, we explore quantum chaos for rotating BTZ in TMG using different approaches. We start with the OTOC, including the subtle periodicity condition, and focus on the quantum chaos in high temperature limit. We compare these chaos parameters with the ones obtained from pole-skipping methods. In section \ref{conclusion} we summarize our findings.

\section{BTZ black holes : a short review}\label{sec:btzreview}

The focus of our work is quantum chaos for rotating BTZ black holes in three dimensional  theories of gravity and we begin with a short review of  basic facts about the rotating BTZ black holes.\footnote{BTZ black holes have been extensively studied in the literature and we refer the reader to \cite{Banados:1992wn,Banados:1992gq,Carlip:1995qv} for details of the geometry and thermodynamics.}
The three dimensional BTZ black holes are vacuum solutions of both Einstein gravity and Topologically Massive Gravity (TMG) with negative cosmological constant.  
BTZ black hole can be obtained from the global AdS$_3$ geometry by an identification of points through the action of a discrete subgroup of $SL(2,\mathbb{R})\times SL(2,\mathbb{R})$, and therefore the geometry of BTZ is locally isomorphic to AdS$_3$.

The metric of a rotating BTZ black hole is given by
\begin{eqnarray}\label{eq:metricbtz}
ds^2=-f(r)dt^2+\frac{dr^2}{f(r)}+r^2\bigg(d\varphi-\frac{r_+ r_-}{\ell  r^2}dt\bigg)^2\,,~\quad f(r) = \frac{(r^2-r_+^2)(r^2-r_-^2)}{\ell^2 r^2}\,.
\end{eqnarray}
Note that $\varphi\sim\varphi+2\pi$ is the angular coordinate and $\ell$ is the AdS radius which will be set to 1 from now on. In most parts of this paper related to pole-skipping, we will assume that the coordinate $\varphi$  is noncompact. This assumption has been widely used in the holographic studies of rotating BTZ, e.g. \cite{Birmingham:2001pj, Son:2002sd, Castro:2014tta}, and is a reasonable assumption that holds in the high temperature limit as we show explicitly in Sec. \ref{sec-otoc}. 

The coordinates $(t,r,\varphi)$ in (\ref{eq:metricbtz}) are often called the Schwarzschild coordinates. 
The metric (\ref{eq:metricbtz}) describes a rotating black hole with inner and outer horizons at $r_-$ and $r_+$ respectively.  
The parameters $r_+$ and $r_-$ are related to the ADM mass $M$, angular momentum $J$, Hawking temperature $T$, and angular potential $\Omega$ of the black hole through
\begin{eqnarray}\label{btzparameters}
M = r_+^2 + r_-^2\,,~~\quad J=2r_+r_-
\,,~~\quad T  =  \frac{r_+^2 - r_-^2}{2 \pi r_+}\,,~~\quad \Omega=\frac{r_-}{r_+}\,.
\end{eqnarray}
As in the case of higher dimenional black holes, BTZ geometry can be extended infinitely by going to the Kruskal coordinates (see \cite{Banados:1992gq} for details and Penrose diagram). The parameter $r_- = 0$ describes a non-rotating black-hole and $r_- = r_+$, is that of an extremal rotating black hole. It is also trivial to verify that for $M=-1, J=0$ we get the empty AdS$_3$. Thus empty AdS$_3$ is seperated from the black hole states by a mass gap.

So far the discussion has been focussed on Einstein gravity. In the presence of a gravitational Chern-Simons term, which we will consider in section \ref{sec:tmg}, the notion of ADM mass and angular momentum changes due to additional surface terms that arise in TMG \cite{Kraus:2005zm}. The appropriate ADM charges in this case are
\be
M(\mu) = M + \frac{J}{\mu}\,,~~~~J(\mu) = J + \frac{M}{\mu}
\ee
where $M$ and $J$ are the charges of Einstein gravity mentioned previously. As expected from the action \eqref{tmgfullaction} in section \ref{sec:tmg}, when the coupling constant $\mu$ is taken to infinity, the theory reduces to Einstein gravity and so does the charges. The conjugate thermodynamic quantities, temperature $T$ and angular velocity $\Omega$, does not depend on the theory since they arise from the regularity of the geometry.

The eternal BTZ black hole is conjectured to be dual to a thermofield double (TFD) state in a CFT with a chemical potential $\Omega$ for angular momentum turned on. Upon decomposing the TFD state to left and right moving sectors, we get two chiral CFTs at temperatures
\begin{eqnarray}
T_L=\frac{T}{1-\Omega}=\frac{r_++r_-}{2\pi}\,, \qquad T_R=\frac{T}{1+\Omega}=\frac{r_+-r_-}{2\pi}\,.
\end{eqnarray}
It is easy to see form \eqref{btzparameters} that $0\leq \Omega \leq 1$, while for a black hole rotating in the opposite sense, we simply have to exchange the left and right modes.
 
 For the near-horizon analysis, it is convenient to use the comoving coordinates ($t, r, \phi$) with 
\be \label{eq:boost}
\phi=\varphi-\frac{r_-}{ r_+} t\,,\ee 
where the metric is
\begin{eqnarray}\label{eq:me-comoving}
ds^2=-f(r)dt^2+\frac{dr^2}{f(r)}+r^2\bigg(\frac{r_-}{r_+}\frac{r^2-r_+^2}{r^2}dt+d\phi\bigg)^2\,.
\end{eqnarray}
In this coordinate the angular velocity at the horizon vanishes $\Omega_H = 0$. 

The dual field theory of (\ref{eq:metricbtz})  lives on the boundary with metric $ds^2=-dt^2+d\varphi^2$ while the dual theory of (\ref{eq:me-comoving}) lives on the spacetime $ds^2=-dt^2+(\Omega dt+d\phi)^2$. These two theories are related by a boost transformation (\ref{eq:boost}).\footnote{The chaos parameters depend on the reference frame \cite{Mezei:2019dfv}.}
The momentum variables in the comoving and Schwarzschild coordinates are related by 
\begin{align}\label{comovingtoSch}
	\begin{split}
		\omega_{\rm sch} = \omega_{\rm cm} + \frac{r_-}{r_+} k_{\rm cm}\,,~~~
		k_{\rm sch} = k_{\rm cm}\,.
	\end{split}
\end{align}

For the purposes of this paper we will need three more coordinate systems, which are collected in appendix \ref{app:threecoords}. In three dimensions, a rotating BTZ black hole metric can always be brought to a non-rotating form by a coordinate transformations as shown in the appendix \ref{app:btzPoincare}. The ingoing Eddington-Finkelstein coordinates and Kruskal coordinates of BTZ are shown in \ref{app:btzEF} and \ref{app:btzKru} respectively. These coordinates are useful for different purposes and after the calculations we will transform back to metric (\ref{eq:metricbtz}) to interpret the physics of quantum chaos. 

\section{Quantum chaos in Einstein gravity}\label{sec:eingravity}

As a warm-up to TMG, we first consider the chaos parameters in Einstein gravity. As shown in  \cite{Stikonas:2018ane,Poojary:2018esz,Jahnke:2019gxr} the rotating BTZ black hole has two sets of Lyapunov exponents $\lambda_{\pm}$, which can be viewed as the high temperature limit of the instantaneous Lyapunov exponent found in \cite{Mezei:2019dfv}, and the butterfly velocities with the speed of light. The average of instantaneous Lyapunov exponent still saturate the chaos bound. 
In this section we shall obtain the same chaos parameters through the phenomenon of pole-skipping, thus confirming the chaos parameters which control the growth of OTOC correlator and scrambling time of mutual information is the same as the pole-skipping point as expected for a black hole.

The action for three dimensional Einstein gravity is
\begin{eqnarray}
S_{\rm EH} = \frac{1}{16\pi G} \int_{\mathcal{M}}d^3x\; \sqrt{-g} \left(R-2\Lambda\right)\,,
\end{eqnarray}
and the equations of motion which follows from the above action is\footnote{$\Lambda$ is related to the AdS radius as  $\Lambda = -1/\ell^2$.}
\begin{eqnarray}\label{eineqn}
R_{ab} - \frac{1}{2}R g_{ab} - g_{ab} = 0\,.
\end{eqnarray}
The BTZ black hole \eqref{eq:metricbtz} is indeed a solution of the Einstein's equation. 

We will present pole-skipping using three different approachs. First, we show that the leading $vv$-component of equations of motion for fluctuations  becomes degenerate at special values in the momentum space, where $v$ is null coordinate of the ingoing Eddington-Finkelstein coordinates as defined in appendix \ref{app:btzEF}. We will also present a strong evidence for the presence of pole-skipping point from the holographic Euclidean correlator and finally we give an explicit CFT calculation of the retarded correlators of $T^{00}$ operator.

Since we are interested in $vv$-component of equations of motion for fluctuations and the stress-tensor correlators, we need linearized metric perturbations around the BTZ background. We expand the metric as $g_{ab} = {\bar g}_{ab}+ h_{ab}$ , where ${\bar g}_{ab}$ is a solution of the vacuum Einstein's equation, and plug into \eqref{eineqn} to obtain the field equations for $h_{ab}$ ,
\begin{align}\label{fluctuation-eom-BTZ}
	\begin{split}
		\; & E_{ab}\equiv\nabla^{c} \nabla_{(a}h_{b)c} - \frac{1}{2} \nabla^{c}\nabla_{c} h_{a b} - \frac{1}{2} \nabla_{a} \nabla_{b} h 
		-  \frac{1}{2} {\bar g}_{a b} \nabla^{c}\nabla^{d} h_{c d}   \\ & \hspace{5cm} 
		+ \frac{1}{2} {\bar g}_{a b} \nabla^{c}\nabla_{c} h
		+ 2 h_{ab} - 2h {\bar g}_{a b} = 0\,,
	\end{split}
\end{align} 
where $h = {\bar g}^{ab}h_{ab}$. The above equations can also be obtained from the second order expansion of the Einstein-Hilbert action around the BTZ background
\begin{align}\label{fluctuation-effective-action}
	\begin{split}
		S  &= \frac{1}{16\pi G}\int d^3x \sqrt{-\bar{g}} \Big[ 2 h_{ab}h^{ab} - h^2    -\frac{1}{2} (\nabla_a h)(\nabla^a h)+ \frac{1}{2} (\nabla_{c}h_{ab})(\nabla^c h^{ab})   \\  & \hspace{2cm}  + (\nabla_a h)(\nabla_b h^{ab}) - (\nabla_b h_{ac})(\nabla^c h^{ab}) \Big]\,.
	\end{split}
\end{align}

 Our first task is to show the ``breakdown" of $E_{vv}=0$ 
 in the ingoing Eddington-Finkelstein coordinates $\{v, r, \phi\}$ near the horizon at special values of $\omega$ and $k$. Let us now consider 
$E_{vv}=0$ in \eqref{fluctuation-eom-BTZ} 
near the horizon $r=r_+$. We expand the solution near the horizon in a series\footnote{As we remarked below (\ref{eq:metricbtz}), we have assumed that the angle coordinate $\phi$ is noncompact, \textit{i.e.}  we are working in the high temperature limit. It would be interesting to include the periodicity of $\phi$ in the  calculations of pole-skipping. }
\begin{eqnarray}\label{fluctuation-ansatz}
h_{ab} = \e^{-i\omega v + i k \phi} (r-r_+)^{\gamma} \sum_{n=0}^{\infty}  \tilde{h}^{(n)}_{ab}(r-r_+)^n\,.
\end{eqnarray}
The $vv$-component of the equations of motion $E_{vv}$ has a second order pole at $r = r_+$ whose coefficient has to be set to zero. This is the leading order equation for $\tilde{h}^{(0)}_{ab}$ and is of the form
\begin{align}\label{eq:Evv}
\begin{split}
\left(2\pi i \omega + 4\pi i \Omega k - k^2 \beta (1 - \Omega^2) \right)\tilde{h}^{(0)}_{vv} &= - (2\pi i-  \beta \omega )(1 - \Omega^2)  \left[ 2 k\tilde{h}^{(0)}_{v \phi} + \omega \tilde{h}^{(0)}_{\phi \phi } \right].
\end{split}
\end{align}

If we now analytically continue the frequencies $(\omega,k)$ to the complex plane, we find that the above equation degenerates (\textit{i.e.} both LHS and RHS becomes zero) at special values of (complex) $\omega$ and $k$
\begin{eqnarray}\label{eq:psincm}
(\omega, k) = \left(\frac{2\pi i}{\beta},\;\pm \frac{2\pi i}{\beta (1 \mp \Omega)}\right)\,.
\end{eqnarray}
It should also be emphasized that other equations of motion are regular at this point. Notice that the Eddington-Finkelstein time coordinate $v$ in the bulk reduces to the usual Schwartzshild time coordinate at the boundary $r\rightarrow \infty$. Thus the pole-skipping points in comoving coordinate (\ref{eq:me-comoving}) take the form (\ref{eq:psincm}).  
 
We can also transform the pole-skipping points (\ref{eq:psincm}) back to the Schwarzschild coordinates (\ref{eq:metricbtz}) using the transformation of frequencies \eqref{comovingtoSch}, and the special points become
\be\label{EinPSpoints}
\left(\frac{2\pi i}{\beta (1\mp \Omega)},\;\pm\frac{2\pi i}{\beta (1\mp \Omega)}\right)\,.
\ee

In the comoving coordinates, from (\ref{eq:psincm}) the Lyapunov exponent saturates the chaos bound while the butterfly velocities satisfy $|v_B| >1 $ and $|v_B|<1$ for the left and right movers respectively. In the Schwarzschild  coordinates, from (\ref{EinPSpoints}) the left mode seemingly violates the chaos bound while the butterfly velocities are always at the speed of light. These results were also obtained from the OTOC calculations of \cite{Stikonas:2018ane,Poojary:2018esz,Jahnke:2019gxr}, however the apparent violation of chaos bound is an artefact of improper boundary conditions along the $
\varphi$ circle, as pointed out in \cite{Mezei:2019dfv}. After the periodicity condition is  taken into account, the OTOC contains only one independent mode and $\lambda_\pm$ are values of the step functions from the instantaneous Lyapunov exponent of  in the high temperature limit \cite{Mezei:2019dfv}. 

We shall provide another evidence for pole-skipping for rotating BTZ black holes in Einstein gravity with the Euclidean two-point function of $T^{00}$ from holography. In three dimensions, Einstein gravity with negative cosmological constant is known to have no propagating gravitational waves and is in fact (classically) equivalent to a Chern-Simons gauge theory with ${\rm SL}(2,\mathbb{R})\times {\rm SL}(2,\mathbb{R})$ gauge group. Therefore the usual holographic procedure of imposing ingoing boundary conditions at the horizon \cite{Son:2002sd} cannot be used to compute the retarded stress-tensor correlators, but we can still find strong evidence for pole-skipping by finding the poles and zeros of the Euclidean correlators because the poles and zeros of retarded correlators will always be contained in the Euclidean correlators. In the subsequent subsection, we calculate the Euclidean correlator following the usual procedure of holographic renormalization calculations. 

\subsection{Euclidean two-point correlator from holography}
\label{sec:e2pchol}
In AdS/CFT, the boundary correlators are obtained from the bulk on-shell action evaluated as a functional of the boundary sources. As is often the case, the on-shell action diverges in the large radius limit and requires adding additional boundary terms to the effective action to render a finite action and stress-tensor. The boundary terms needed for Einstein gravity with Dirichlet boundary conditions was found in \cite{Balasubramanian:1999re,deHaro:2000vlm} and the  renormalized action is given by
\begin{align}\label{reneinaction}
S_{\rm ren.} &= \frac{1}{16\pi G} \int_{\mathcal{M}}d^3x\; \sqrt{g} \left(R+2\right) + \frac{1}{8\pi G} \int_{\partial \mathcal{M}} d^2x\;\sqrt{\gamma} \;( K-1)
\end{align}
where $K$ is the trace of extrinsic curvature. To proceed, we will work with the metric \eqref{poincarebhmetric} and impose the radial gauge $h_{\rho a} = 0$ condition on the fluctuations. In this coordinates, the boundary CFT lives in the Minkowski space-time with metric $-dT^2 + dX^2$ and the bulk fields $h_{ij}$ are dual to the stress-tensor of the boundary CFT at finite temperature.

In order to calculate the (Euclidean) on-shell action we first Wick rotate the Poincare time coordinates of \eqref{poincarebhmetric} by  $T\rightarrow i\tau$
\begin{eqnarray}\label{poincareebhmetric}
ds^2 = \frac{d\rho^2}{4\rho^2}  + \frac{1}{\rho} \Big[(1-\rho)^2 d\tau^2 + (1+\rho)^2 dX^2 \Big]
\end{eqnarray}
where $\tau$ gets identified as $\tau \sim \tau + \pi$ to avoid conical singularity at $\rho =1$. We can now solve the field equations using the plane wave ansatz
\begin{eqnarray}\label{PoincareEinEucSoln}
h_{ij}(\rho,\tau,X) = \e^{i \omega_E \tau + i k X}\;\frac{1}{\rho}\;\tilde{h}_{ij}(\rho)\,,~~~\quad \tilde{h}_{ij}(\rho) = \tilde{h}^{(0)}_{ij} + \rho \tilde{h}^{(1)}_{ij} + \rho^2 \tilde{h}^{(2)}_{ij}
\end{eqnarray}
where $\tilde{h}_{ij}$ is determined as an expansion near the boundary. The leading coefficient $\tilde{h}^{(0)}_{ij}$ can be identified with the source term for the stress tensor. In three dimensional Einstein gravity, it is a well-known fact that the expansion terminates at $\rho^2$ in the absence of matter fields and the coefficients $\tilde{h}^{(1)}_{ij}$ and $\tilde{h}^{(2)}_{ij}$ can be determined in terms of $\tilde{h}^{(0)}_{ij}$ by solving the equations of motion order by order in $\rho$, and explicit expressions for the coefficients in (\ref{PoincareEinEucSoln}) can be found in appendix \ref{app:coe3.13}.

The effective action for the fluctuations \eqref{fluctuation-effective-action}, up on substituting  the equations of motion \eqref{fluctuation-eom-BTZ}, reduce to a surface integral over the boundary
\begin{align}
\begin{split}
S_{\rm o.s.}  =\frac{1}{16\pi G} \int_{\partial\mathcal{M}} d^2 x \sqrt{-\gamma} n^a \bigg[ -\frac{1}{2} h \nabla_a h + \frac{1}{2} h \nabla^b h_{ab} + \frac{1}{2} h_{ab} \nabla^b h  
- h^{bc}\nabla_b h_{ac} + \frac{1}{2} h^{bc}\nabla_a h_{bc} \bigg] 
\end{split}
\end{align}
where $\gamma_{ab}$ is the induced metric at the boundary and $n^a$ is the outward pointing normal vector to the boundary hypersurface. 
By expanding the renormalized action \eqref{reneinaction} to second order in fluctuations, we see that the counterterms which render finite on-shell action takes the form 
\begin{eqnarray}
S_{\text{c.t.}} = -\frac{1}{16\pi G}\int_{\partial \mathcal{M}} \sqrt{-\gamma} \big(h_{ab}h^{ab} - \frac{1}{2}h^2 \big)\,.
\end{eqnarray}

The Green's function can be obtained by functionally differentiating the renormalized on-shell action w.r.t. $\tilde{h}^{(0)}_{\tau\tau}$ twice\footnote{This should be thought of as a schematic representation of the correlation function. A proper treatment includes subtracting the contact terms also from the on-shell action.}
\begin{eqnarray}
\label{eq:euc2pt}
\bra T^{\tau\tau}(\omega_E,k)T^{\tau\tau}(-\omega_E,-k)\ket \propto \frac{\delta^2 S_{\text {ren.}}}{\delta \tilde{h}^{(0)}_{\tau\tau}\delta \tilde{h}^{*(0)}_{\tau\tau}}  = \frac{k^2(4+k^2)}{2(\omega_E^2 + k^2)}\,.
\end{eqnarray}
From the above correlator, we see that $\omega_E = \pm 2$, $k = \pm 2i$ are indeed special points where the numerator and the dinominator vanishes. We can now Wick rotate $\omega_E \rightarrow i \omega$ and transform the points back to the momentum space for metric (\ref{eq:metricbtz})  using \eqref{momentum-transformation} obtaining, 
\begin{eqnarray}
\left(\frac{2\pi i}{\beta (1\mp \Omega)},\;\pm\frac{2\pi i}{\beta (1\mp \Omega)}\right)\,.
\end{eqnarray}
These are the same points we found in \eqref{EinPSpoints} and they correspond to $\omega = 2i$. For $\omega= -2i$, we find ${\rm Im}(\omega_\text{BTZ})<0$ and in Schwarzschild coordinates, they correspond to
$
\big(-\frac{2\pi i}{\beta (1\pm \Omega)},\;\pm\frac{2\pi i}{\beta (1\pm \Omega)}\big).
$
These points are exponentially suppressed and therefore do not contribute to the growth of OTOC correlators. 

\subsection{Retarded Green's function from CFT}
\label{sec:rgfCFT}
The fact that there are special points where retarded Green's function is undefined can also be understood from explicit CFT calculations. Pole-skipping from Euclidean correlators has been studied in \cite{Haehl:2018izb} and we shall focus on the retarded Green's function in this subsection. 
In two dimensions, conformal symmetry is powerful enough to allows us to calculate the retarded two-point function for $T^{00}$ operator. This calculation of retarded correlator is closly adapted from \cite{Birmingham:2001pj,Sabella-Garnier:2019tsi}.

We begin by recalling the holomorphic and anti-holomorphic two point functions of stress tensor on the complex plane
\begin{eqnarray}
\langle T(z) T(z')\rangle=\frac{c_L/2}{(z-z')^4}\,,~~~~\langle \bar{T}(\bar{z})\bar{T}(\bar{z}')\rangle=
\frac{c_R/2}{(\bar{z}-\bar{z}')^4}
\end{eqnarray}
where $c_L$ and $c_R$ are the left/right central charges. In a CFT dual to Einstein gravity, $c_L=c_R=\frac{3}{2G}$. However we shall carry forward the notation because later in TMG we encounter dual CFT with unequal central charges. 

Our first task is to map the above plane correlator to a cylinder with complex coordinate $(w,\bar{w})$ where $w=\sigma+i\tau$ in order to match the results with the BTZ black hole. The conformal transformation that maps the plane to the cylinder is 
\begin{eqnarray}\label{planetocycl}
z=\e^{\frac{2\pi}{\beta_L}w},\quad\bar{z}=\e^{\frac{2\pi}{\beta_R}\bar{w}}.
\end{eqnarray}
The transformation \eqref{planetocycl} is not an element of the global conformal group, therefore the stress-tensor transforms with a non-zero Schwarzian derivative and we get\footnote{One can perform a Fourier transformation of (\ref{eq:euc2pt}) to obtain this form in the coordinate space if we also include the right moving modes, up to the Casimir terms. 
}
\begin{eqnarray}
\bra T(w)T(w')\ket_{\rm cycl} = \left(\frac{\pi^2 c_L}{6\beta_L} \right)^2+ \frac{c_L}{32}\left(\frac{2\pi}{\beta_L}\right)^4 \frac{1}{\sinh^4\left[\frac{\pi}{\beta_L}(w-w')\right]}\,.
\end{eqnarray}
The transformation of the right moving sector is completely analogous except for the replacement $\beta_L\rightarrow\beta_R, w\rightarrow \bar{w}$. Notice that we have used the fact that the CFT dual to a rotating BTZ black hole has different temperatures for the left and right moving sectors. The two temperatures are related via
\begin{eqnarray}
\label{eq:tem}
\beta_L = \beta (1-\Omega)\,,\quad \beta_R = \beta(1+\Omega)\,.
\end{eqnarray}

Since we are interested in correlators of energy density \textit{i.e.} $\langle T^{00} (t,\sigma)T^{00}(0,0) \rangle$, we have to transform from the holomorphic/anti-holomorphic coordinates to the Cartesian coordinates  where the correlators are related by
\begin{eqnarray}
\langle T^{00}(\tau,\sigma) T^{00}(0,0) \rangle = \langle T(\sigma+i\tau)T(0) \rangle_{\rm cycl} + \langle \bar{T}(\sigma-i\tau)\bar{T}(0) \rangle_{\rm cycl}\,.
\end{eqnarray}
To obtain Lorentzian correlators, we have to analytically continue the above correlator by taking $\tau = i t\pm \epsilon$. The appropriate sign of $\epsilon$ is decided by the ordering of operators inside the real-time correlators. The retarded two-point function is given by 
\begin{eqnarray}
G^{R}_{T^{00}T^{00}}(t,\sigma) = i\theta(t)\bra \left[T^{00}(t,\sigma),T^{00}(0,0)\right]\ket = i\theta(t)G(t,\sigma)\,.
\end{eqnarray}
In the above expression $\left[\;,\;\right]$ denotes the commutator of two operators and $G(t,\sigma) = G_+(t,\sigma)-G_-(t,\sigma)$. The Casimir term drops out in the expression for $G(t,\sigma)$ due to the commutator. For $G_+$ the $t$ contour has to be deformed by $t-i\epsilon$ whereas for $G_-$ the deformation is $t+i\epsilon$
\begin{align}
\begin{split}
G_{\pm} = \frac{c_L}{2}\left(\frac{\pi/\beta_L}{\sinh\left[\frac{\pi}{\beta_L}(\sigma-t \pm i\epsilon)\right]} \right)^4 + \frac{c_R}{2}\left(\frac{\pi/\beta_R}{\sinh\left[\frac{\pi}{\beta_R}(\sigma + t \mp i\epsilon)\right]} \right)^4 \,.
\end{split}
\end{align}

The Fourier transform of $G(t,\sigma)$ is 
\begin{align}\label{GFourierTrans}
\begin{split}
G(\omega,k) &= \frac{c_L}{6}  \left(\frac{2\pi}{\beta_L}\right)^3 \delta\left(\frac{k-\omega}{2}\right) \sinh\left[\frac{\beta_L}{2}\left(\frac{k+\omega}{2}\right) \right] \left|\Gamma\left(2+\frac{i \beta_L}{2\pi}\left(\frac{k+\omega}{2}\right)\right) \right|^2  \\ &-   \frac{c_R}{6} \left(\frac{2\pi}{\beta_R}\right)^3 \delta\left( \frac{k+\omega}{2} \right) \sinh\left[\frac{\beta_R}{2}\left(\frac{k-\omega}{2}\right) \right] \left|\Gamma\left(2+\frac{i \beta_R}{2\pi}\left(\frac{k-\omega}{2}\right)\right) \right|^2\,.
\end{split}
\end{align}
To evaluate the Fourier integral, we have made a change of variable to $x^{\pm} = \sigma \pm t$ and used the identity \cite{Gubser:1997cm,Bredberg:2009pv}
\begin{eqnarray}
\int dx\; \e^{ikx}\left(\frac{\pi}{\beta\sinh\left[\frac{\pi x\pm i\epsilon}{\beta}\right]}\right)^4
=\frac{1}{6}\left(\frac{2\pi}{\beta} \right)^3\e^{\mp \frac{k\beta}{2}}
\left| \Gamma \left(2+\frac{ik\beta}{2\pi} \right)\right|^2\,.
\end{eqnarray}
Finally, the Fourier transform of the full retarded correlator can be obtained by a convolution integral of \eqref{GFourierTrans} with the Fourier coefficient of Heaviside step function and we get
\begin{align}\label{retT00T00corr}
\begin{split}
G_R (\omega,k) 
&= \frac{c_L}{6} \left(\frac{2\pi}{\beta_L}\right)^3\left(\frac{2i}{\omega-k}+\pi\delta\big(\frac{k-\omega}{2}\big)\right)\sinh\left[\frac{\beta_L k}{2} \right]\left|\Gamma\left(2+\frac{i \beta_L}{2\pi}k \right) \right|^2\\
&~~~
-\frac{c_R}{6} \left(\frac{2\pi}{\beta_R}\right)^3\left(\frac{2i}{\omega+k}+\pi\delta\big(\frac{k+\omega}{2}\big)\right)
\sinh\left[\frac{\beta_R k}{2} \right]\left|\Gamma\left(2+\frac{i \beta_R}{2\pi}k \right) \right|^2\,.
\end{split}
\end{align}

We are finally ready to show the presence of special points in the analytic structure of retarded correlators. In the first term in (\ref{retT00T00corr}), the hyperbolic sine function 
has simple zeros at discrete values along the imaginary $k$ axis 
\begin{eqnarray}
k = \frac{2\pi i m}{\beta_L}\,, \qquad m\in \ZZ\,.
\end{eqnarray}
The gamma function contributes two branch of simple  poles which also lie on the imaginary $k$ axis
\begin{eqnarray}
k = \pm 
\frac{2\pi i}{\beta_L} \left( p+2\right)\,, \qquad p=0,1,2,\cdots\,.
\end{eqnarray}
Therefore the combined function $\sinh\left[\frac{\beta_L k}{2} \right] \left|\Gamma\left(2+\frac{i \beta_L}{2\pi}k \right) \right|^2$ only has zeros at $k=0, \pm\frac{2\pi i}{\beta_L}$ while no poles. 
From the first term in (\ref{retT00T00corr}) the prefactor $1/(\omega - k)$ restricts the poles on the $\omega$ plane to be on $\omega = k$. The pole-skipping points for first term in (\ref{retT00T00corr}) are at
\begin{eqnarray}
\label{eq:term1}
(\omega,k) =  \left(\pm \frac{2\pi i}{\beta (1-\Omega)},~\pm \frac{2\pi i}{\beta (1-\Omega)}\right)\,,
\end{eqnarray}
where we have used (\ref{eq:tem}). 

Similar arguments can also be made 
for the second term in (\ref{retT00T00corr}) and in this case 
the special pole-skipping points for  are given by
\begin{eqnarray}
\label{eq:term2}
 (\omega,k) =  \left(\pm \frac{2\pi i}{\beta (1+\Omega)},~\mp \frac{2\pi i}{\beta (1+\Omega)}\right)\,.
\end{eqnarray}
As we found from the Euclidean correlator, there are two set of pole-skipping points corresponding to the positive and negative values of ${\rm Im}(\omega_\text{BTZ})$. It is the positive pairs which is responsible for the exponential growth of OTOC and they match well with the pole-skipping points from the near horizon equations of motion (\ref{EinPSpoints}). Note that to calculate the pole-skipping in this subsection we have not made any special assumptions on the underlying CFT. In other words, for non-chaotic or integrable CFT, there also exists pole-skipping. Therefore in generic non-chaotic system, pole-skipping and 
OTOC do not necessary to give the same result of chaos parameter. Nevertheless for the case of rotating BTZ in Einstein gravity, at high temperature OTOC and pole-skipping give the same chaos parameters. 

\section{Quantum chaos in topologically massive gravity}\label{sec:tmg}
Topologically Massive Gravity (TMG) is a modification of Einstein gravity with the addition of a gravitational Chern-Simons term and was first introduced in \cite{Deser:1981wh,Deser:1982vy}. For a more contemporary study of TMG, see \cite{Li:2008dq, Castro:2014tta, Kraus:2005zm,  Skenderis:2009nt} and references therein.
The action for the three dimensional TMG consists of the gravitational Chern-Simons term in addition to the Einstein-Hilbert term \cite{Deser:1981wh,Deser:1982vy}\footnote{Convention: $\varepsilon_{abc}=\sqrt{-g}\epsilon_{abc}$ with $\epsilon_{012}=1$.  
	In this paper we shall focus on the regime $\mu>0$.
}
\begin{eqnarray}\label{tmgfullaction}
S=\frac{1}{16\pi G}\int d^3 x \sqrt{-g}\left(R+2+\frac{1}{2\mu} \varepsilon^{abc}\Gamma^d_{~ae}\left(\partial_b\Gamma^e_{~cd}+\frac{2}{3}\Gamma^{e}_{\;bf}\Gamma^f_{\;cd}\right) \right)\,.
\end{eqnarray}

The equations of motion that follows from the action is given by
\begin{eqnarray}\label{tmgeom}
R_{ab}-\frac{1}{2}g_{ab}\left(R+2\right)+\frac{1}{\mu}C_{ab} =0
\end{eqnarray}
where $C_{ab}$ is the Cotton tensor
\begin{eqnarray} 
C_{ab}=\varepsilon_{a}^{~cd} \nabla_c S_{d b}\,, \quad S_{ab}=R_{ab}-\frac{1}{4}g_{ab}R\,.
\end{eqnarray}
We can immediately see that the equations of motion are third order in derivatives, so in order to solve the field equation we need to specify the metric and its derivatives at the boundary. We also notice that substituting $R_{ab}=-2g_{ab}$ solves the field equations, so solutions of Einstein gravity are solutions of TMG as well, which is true of BTZ black hole in particular.

Unlike CFT dual to Einstein gravity, for asymptotically AdS$_3$ solutions in TMG the left and right CFTs carry different central charge
\begin{eqnarray}\label{eq:TMGcc}
(c_L,c_R) = \frac{3\ell}{2G}\left(1-\frac{1}{\mu},1+\frac{1}{\mu}\right)\,,
\end{eqnarray}
leading to a gravitational anomaly in the dual CFT. 

In the following parts of this section, we shall calculate the  chaos parameters for rotating BTZ in TMG using different methods. We start from the shock wave calculations imposing the periodicity condition on the phase factors. We then study the pole-skipping phenomenon as well from different perspectives, including the near horizon equation of motion, the retarded Green's function from holography and from CFT. 

\subsection{Out-of-time-order correlator from shock waves}
\label{sec-otoc}
 
The out-of-time-order correlator (OTOC) can be calculated in holography by considering the scattering amplitude of bulk  
particles dual to $V$ and $W$ operators in a black hole geometry \cite{Shenker:2013pqa, Shenker:2014cwa}. 
In the elastic eikonal gravity approximation, the phase shift in the amplitude is determined by the classical on-shell action on the gravity side, and is proportional to the profile of a shock wave localized on the horizon $U=0$ in the Kruskal coordinate (see appendix \ref{app:btzKru}). The OTOC calculation for the rotating BTZ black hole in Einstein gravity was carried out in \cite{Jahnke:2019gxr} and we refer this paper for the details of the calculation. In the TMG case, the only distinction comes from the eikonal phase factor which depends on the gravitational theory. 

For TMG, we start with a perturbation of form 
$
\delta T_{UU}=-\frac{1}{8\pi G r_+^2}\delta(U)\delta(\phi)
$
which generates $ ds^2\to ds^2+\frac{2\ell^2}{(1+UV)^2}\delta(U)h(\phi)dU^2$, and  
from the $UU$-component of the equations of motion 
in Kruskal coordinates (\ref{eq:btzKruskal}) we have 
\begin{eqnarray}
\begin{split}\label{eq:swequation}
h'''(\phi) + (r_+ \mu -3r_-)h''(\phi) &+ (3r_-^2 - r_+^2-2r_+r_-\mu)h'(\phi) \\ &- (r_-^3 - r_+^2 r_- -\mu r_+r_-^2 +  \mu r_+^3)h(\phi)=\#\delta(\phi)\,. 
\end{split}
\end{eqnarray}
The most general solution to the above equation is given by\footnote{Note that here we absorbed the step-functions into the coefficients $c_1, c_2$ and $c_3$.}
\begin{eqnarray}
\label{eq:swsolution}
h(\phi) = c_1 \; e^{-\frac{2\pi\phi}{\beta(1+\Omega)}} +c_2 \; e^{ \frac{2\pi\phi}{\beta(1-\Omega)}}+c_3\;e^{\frac{2\pi(\Omega-\mu)\phi}{\beta(1-\Omega^2)}} \,.
\end{eqnarray}
Note that the first two terms in (\ref{eq:swsolution}) also appeared in Einstein gravity \cite{Jahnke:2019gxr} while the third term is a new mode which is related to the fact that we have more degrees of freedom for the fluctuations in TMG.  
When $\phi$ is noncompact, one should impose the condition $|\phi|\to\infty$, $h(\phi)\to 0$ to determine the coefficients \cite{Shenker:2014cwa}. The final solution takes the following form, with $c_0$ defined as  $c_0=\frac{\#}{2r_+^2 (1-\mu^2)}$, and for $\mu\neq 1$ 
\begin{itemize}

\item When $\mu > \Omega$
\begin{eqnarray}
\label{eq:planarsw1}
h(\phi)=\begin{cases}
 c_0\bigg[(1+\mu)\;e^{ \frac{-2\pi\phi}{\beta(1+\Omega)}} -2\;e^{\frac{2\pi(\Omega-\mu)\phi}{\beta(1-\Omega^2)}} \bigg] \,, \quad\quad  &\text{if} \;\;  \phi>0\\
- c_0 (1-\mu)\; e^{\frac{2\pi\phi}{\beta(1-\Omega)}}  \,, \quad\quad  &\text{if} \;\; \phi<0 \\
	\end{cases}\,
\end{eqnarray}

\item When $\Omega >\mu$
\begin{eqnarray}
\label{eq:planarsw2}
h(\phi)=\begin{cases}
c_0 (1+\mu)\; e^{-\frac{2\pi\phi}{\beta(1+\Omega)}}  \,, \quad\quad  &\text{if} \;\; \phi>0 \\
 -c_0\bigg[(1-\mu)\;e^{ \frac{2\pi\phi}{\beta(1-\Omega)}} -2\;e^{\frac{2\pi(\Omega-\mu)\phi}{\beta(1-\Omega^2)}} \bigg] \,, \quad \quad &\text{if} \;\;  \phi<0\\
	\end{cases}\,
	\end{eqnarray}

\item When $\Omega=\mu$
\begin{eqnarray}
\label{eq:planarsw3}
h(\phi)=\begin{cases}
\frac{\#}{r_+-r_-}\; e^{-\frac{2\pi\phi}{\beta(1+\Omega)}}  \,, \quad\quad  &\text{if} \;\; \phi>0 \\
-\frac{\#}{r_-^2-r_+^2}-
\frac{\#}{2r_+(r_++r_-)}\;e^{ \frac{2\pi\phi}{\beta(1-\Omega)}} \,, \quad \quad &\text{if} \;\;  \phi<0\\
	\end{cases}\,
	\end{eqnarray}
 
 \end{itemize}
When $\mu\to\infty$, $c_0\sim \mathcal{O}(1/\mu) $ and 
(\ref{eq:planarsw1}) should give rise to the results in \cite{Jahnke:2019gxr}. When $\Omega=0$, the profile of the shock wave reduced to that for the non-rotating BTZ in TMG \cite{Alishahiha:2016cjk}. Also note that when $\Omega=\mu$ the last term in (\ref{eq:swsolution}) becomes a constant  and in this case we cannot have a solution with $h(|\phi|\to\infty)\to 0$. We will not consider this case in the following.

When $\mu=1$, the third term in (\ref{eq:swsolution}) is exactly the same as the first term and the system will have a new solution of form $\phi \;e^{-\frac{2\pi\phi}{\beta(1+\Omega)}}$, which is quite similar to what happens for  the linear graviton solution in TMG \cite{Grumiller:2008qz}.\footnote{When $\mu=1$, this new  linear graviton solution should be excluded under the Brown-Henneaux boundary condition.} 
We have 
\begin{eqnarray}
\label{eq:planarsw4}
h(\phi)=\begin{cases}
\frac{\#}{4r_+}\; (1+2 r_+\phi)\:e^{-\frac{2\pi\phi}{\beta(1+\Omega)}} \,, \quad\quad  &\text{if} \;\; \phi>0 \\
\frac{\#}{4r_+^2}\;e^{ \frac{2\pi\phi}{\beta(1-\Omega)}} \,, \quad \quad &\text{if} \;\;  \phi<0\\
	\end{cases}\,.
	\end{eqnarray}
	The OTOC at this special point $\mu=1$ will be studied in subsection \ref{sec:cp}. 

If we substitute the solutions (\ref{eq:planarsw1}, \ref{eq:planarsw2}) and (\ref{eq:planarsw4}) in the OTOC,  
the growth of OTOC, $g(t,\varphi)$  
is governed by three set of Lyapunov exponents and corresponding velocities
\begin{eqnarray}\label{otoc-formula}
g(t,\varphi) = 1 - \varepsilon_{VW}\; e^{\frac{2\pi}{\beta}t}h(\Omega t - \varphi) \,
\end{eqnarray}
where $\varphi$ is the angular coordinate in (\ref{eq:metricbtz}) and we restrict to $t\gg \beta$. 

The Lyapunov exponents from the above naive observation are of form   
\begin{eqnarray}\label{eq:lya-tmg}
 \lambda_{\pm} = \frac{2\pi}{\beta(1\mp \Omega)}\,, \quad \lambda_m = \frac{2\pi(1-\mu \Omega)}{\beta(1-\Omega^2)}\,,
 \end{eqnarray}
with the corresponding butterfly velocities of the Lyapunov exponents 
 \be\label{eq:vel-tmg}
 v_\pm=\pm 1 \,,~~~v_m=\frac{1-\mu\Omega}{\Omega-\mu}\,.
 \ee
When $\Omega=0$, the above formulae reduced to the one found in \cite{Alishahiha:2016cjk} where the butterfly velocities for non-rotating BTZ in TMG has been studied using shock wave approach. 

However, 
the above formulae is true for the case that $\varphi$ is noncompact which is true for high temperature case.  As shown in \cite{Mezei:2019dfv}, the correct interpretation of the OTOC requires  considering the periodicity of $h(\phi)$. We will show in the high temperature limit the Lyapunov exponents can be (partly) reproduced after considering this subtlety.

We first impose the periodicity of $\phi$ by $h(\phi)\rightarrow h(\phi \; {\rm mod}\;2\pi)$. This can be done by replacing the $\delta(\phi)$ in (\ref{eq:swequation}) by $\sum_{n=-\infty}^{\infty}\delta(\phi-2\pi n)$ following {\it e.g.}  \cite{Sfetsos:1994xa}. Therefore, when we restrict $\phi\in [0, 2\pi)$
 the solutions should be 

\begin{itemize}

\item When $\mu > \Omega$
\begin{eqnarray}
\begin{split}
h(\phi)&=
 c_0\sum_{n=-\infty}^{0}  \left[(1+\mu)\;e^{ \frac{-2\pi(\phi-2\pi n)}{\beta(1+\Omega)}} -2\;e^{\frac{2\pi(\Omega-\mu)(\phi-2\pi n)}{\beta(1-\Omega^2)}} \right] 
- c_0 (1-\mu)\; \sum_{n=1}^\infty e^{\frac{2\pi(\phi-2\pi n)}{\beta(1-\Omega)}}   \\
&=\frac{c_0(1+\mu)}{1-e^{-\frac{4\pi^2}{\beta (1+\Omega)}}}e^{ \frac{-2\pi\phi}{\beta(1+\Omega)}} 
-\frac{c_0(1-\mu)}{e^{\frac{4\pi^2}{\beta (1-\Omega)}}-1}e^{\frac{2\pi\phi}{\beta(1-\Omega)}} 
-\frac{2c_0}{1-e^{\frac{4\pi^2 (\Omega-\mu)}{\beta(1-\Omega^2)}}}e^{\frac{2\pi(\Omega-\mu)\phi}{\beta(1-\Omega^2)}}
\end{split}
\end{eqnarray}

\item When $\Omega >\mu $
\begin{eqnarray}
\begin{split}
h(\phi)&=
c_0 (1+\mu)\; \sum_{n=-\infty}^{0} e^{-\frac{2\pi(\phi-2\pi n)}{\beta(1+\Omega)}}  
 -c_0 \sum_{n=1}^\infty\left[(1-\mu)\;e^{ \frac{2\pi(\phi-2\pi n)}{\beta(1-\Omega)}} -2\;e^{\frac{2\pi(\Omega-\mu)(\phi-2\pi n)}{\beta(1-\Omega^2)}} \right] \\
 &=\frac{c_0(1+\mu)}{1-e^{-\frac{4\pi^2}{\beta(1+\Omega)}}}e^{-\frac{2\pi \phi}{\beta(1+\Omega)}}
-\frac{c_0(1-\mu)}{e^{\frac{4\pi^2}{\beta (1-\Omega)}}-1}e^{\frac{2\pi\phi}{\beta(1-\Omega)}} 
-\frac{2c_0}{1-e^{\frac{4\pi^2 (\Omega-\mu)}{\beta(1-\Omega^2)}}}e^{\frac{2\pi(\Omega-\mu)\phi}{\beta(1-\Omega^2)}}
 \end{split}
\end{eqnarray}	

\end{itemize}

Note that in the above formula, we have restricted $\phi\in [0, 2\pi)$. In this parametrization, we can readily see that in the limit $\mu \rightarrow \infty$, using $c_0\sim \mathcal{O}(1/\mu)$ the coefficients reduce to that of Einstein gravity in \cite{Mezei:2019dfv}.  
Interestingly, we find that in cases whenever $\mu\neq \Omega$, we have the unique result for the profile of the shock wave $h(\phi)$\footnote{We have set the coefficient $c_0$ to be $1$.}
\be
\label{eq:swcompact}
h(\phi)
 =\frac{1+\mu}{1-e^{-\frac{4\pi^2}{\beta(1+\Omega)}}}\;e^{-\frac{2\pi \phi}{\beta(1+\Omega)}}
-\frac{1-\mu}{e^{\frac{4\pi^2}{\beta (1-\Omega)}}-1}\;e^{\frac{2\pi\phi}{\beta(1-\Omega)}} 
-\frac{2}{1-e^{\frac{4\pi^2 (\Omega-\mu)}{\beta(1-\Omega^2)}}}\;e^{\frac{2\pi(\Omega-\mu)\phi}{\beta(1-\Omega^2)}}\,.
\ee

The OTOC (\ref{otoc-formula}) can now be written as
\be
\label{eq:otoc-gtphi}
g(t,\varphi)=1-\varepsilon_{VW} e^{\frac{2\pi}{\beta}t} h(\Omega t-\varphi) 
\ee
where $h$ takes the form of (\ref{eq:swcompact}). 

It would be interesting to perform the OTOC calculations in CFT following \cite{Roberts:2014ifa}. In the CFT calculation one needs to include the additional contributions from the conformal blocks related to the operator which is dual to the massive graviton in TMG. From field theory the ratio between the coefficients in front of the terms  from vacuum conformal block and the new conformal block should be operator independent  in large $c$ limit due to the unique value of OTOC obtained from holography.\footnote{Note that similar results have been found in \cite{Alishahiha:2016cjk}.} 

The periodicity of the comoving angular coordinates $\phi$ translates to the periodicities of $t \sim t+ \frac{2\pi}{\Omega}$ and $\varphi \sim \varphi + 2\pi$ and it ensures the modulation of the phase factor. 
One can now study the instantaneous Lyapunov exponent  \cite{Mezei:2019dfv} which is defined as
\begin{eqnarray}\label{lam-inst}
\lambda_\text{inst.}(t) =\frac{2\pi}{\beta}+\frac{\partial_t h(\Omega t)}{h(\Omega t)}\,.
\end{eqnarray} 
In the following we will study the high temperature limit of this instantaneous Lyapunov exponent.

\subsubsection{High temperature limit of instantaneous Lyapunov exponents}
\label{sec:instan}

In order to better understand 
the connection between Lyapunov exponents of the OTOC and pole-skipping, it is instructive to look at the high temperature, $\beta \rightarrow 0$ limit. We first consider the $\beta \rightarrow 0$ limit of $\lambda_\text{inst.}$. Due to the periodicity $2\pi/\Omega$ of $t$, it is enough to restrict to the period  $0 \leq t \leq 2\pi/\Omega$\footnote{Note that \eqref{otoc-formula} is valid in the regime $t_d\ll t < t_*$ where $t_d$ is the dissipation time and $t_*$ is the scrambling time. In the high temperature limit, we have $t_d\sim\beta$  and the ensuing discussion is within  this regime.} and we shall comment on the high temperature behavior of $\lambda_\text{inst.}$ below.
\begin{itemize}
\item When $\mu> 1$, $\lambda_\text{inst.}$  
takes following piece-wise form,
\begin{eqnarray}
\lambda_\text{inst.}=\begin{cases}
\lambda_-  \,, \quad &\text{if} \;\;  t \in\big[0, ~ \frac{\pi (1+\Omega)}{\Omega}\big)\\
  \lambda_+\,, \quad &\text{if} \;\;  t \in\big[ \frac{\pi (1+\Omega)}{\Omega}, ~
  \frac{2\pi}{\Omega}\big)\\
	\end{cases}\,.~~
	\end{eqnarray}
In a periodic time of $\frac{2\pi}{\Omega}$, we have the average $\langle \lambda_\text{inst.} \rangle=\frac{2\pi}{\beta}$. 
Note that only two of the Lyapunov exponents in (\ref{eq:lya-tmg}) $\lambda_\pm$ are reproduced in the high temperature limit.  

\item  When $\mu<1$, the results depend on the relative values between $\Omega$ and $\mu$.
If $ \Omega <\mu$, we have
\begin{eqnarray}
\lambda_\text{inst.}=\begin{cases}
\lambda_m  \,, \quad &\text{if} \;\;  t \in\big[0, ~ \frac{2\pi(1+\Omega)}{\Omega(1+\mu)}\big)\\
   \lambda_+\,, \quad &\text{if} \;\;  t \in\big[ \frac{2\pi(1+\Omega)}{\Omega(1+\mu)}, ~
   \frac{2\pi}{\Omega}\big)\\
	\end{cases}\,,~~~
\nn \end{eqnarray}
thus we have average $\langle \lambda_\text{inst.}\rangle=\frac{2\pi}{\beta}$. 

While in the case $\mu <\Omega$, we have 
\begin{eqnarray}
\lambda_\text{inst.}=\begin{cases}
\lambda_-  \,, \quad &\text{if} \;\;  t \in\big[0, ~ \frac{2\pi(\Omega-\mu)}{\Omega(1-\mu)}\big)\\
  \lambda_m\,, \quad &\text{if} \;\;  t \in\big[ \frac{2\pi(\Omega-\mu)}{\Omega(1-\mu)}, ~
  \frac{2\pi}{\Omega}\big)\\	\end{cases}\,.~~~
	\nn
		\end{eqnarray} 	
In this case, we also have average $\langle \lambda_\text{inst.}\rangle= \frac{2\pi}{\beta} $. 	
 \end{itemize}

In the above formulae  we have used $\lambda_\pm, \lambda_m$ defined in the equation (\ref{eq:lya-tmg}).
One can see that if $\mu <  1$, the instantaneous Lyapunov exponents from $\beta\to 0$ limit produce all three Lyapunov exponents $\lambda_\pm, \lambda_m$ and the average 
Lyapunov exponent saturates the chaos bound. While for the regime $\mu > 1$, the $\beta\to 0$ can only produce two of the Lyapunov exponents. We have numerically checked that for any temperature the average of instantaneous Lyapunov exponents is $\frac{2\pi}{\beta}$ and saturates the MSS chaos bound.
 
\subsubsection{Lyapunov Exponents and Butterfly Velocities from OTOC}
\label{sec:lyexotoc}

By including the $\varphi$ dependence and taking the $\beta \rightarrow 0$ limit of the OTOC $g(t,\varphi)$, we can get the information about the butterfly velocities of various modes. We consider the situation where $|\Omega t -\varphi|\ll1$ where the angular coordinate can essentially be regarded as a flat direction. We have to distinguish into cases
in the same way as the previous calculations of the instantaneous Lyapunov exponents.  
\begin{itemize}

\item For $\mu >1$, the OTOC is given by 
\begin{eqnarray}
	g(t,\varphi) = 1 -\varepsilon_{VW} \begin{cases}
	(\mu-1)\;
	e^{\lambda_+\left(t-\varphi\right)}  \,, \quad & \text{if} \;\; \Omega t <\varphi\, \\
	(1+\mu)\; e^{\lambda_-(t+\varphi)} \,, \quad & \text{if} \;\; \Omega t \geq \varphi \,\;\;
	\end{cases}\,.
	\end{eqnarray}

Thus the Lyapunov exponent is $\lambda_-$  and corresponding butterfly velocity $v_-=-1$. The chaos bound proposed in \cite{Maldacena:2015waa} is always saturated and one can view the dual system as a non-maximally chaotic system. The butterfly velocities are always equal to 
speed of light. 

The speed depend Lyapunov exponent is 
\begin{eqnarray}
\label{eq:vdle1}
\lambda(v) = \frac{2\pi}{\beta} \begin{cases}
	\frac{1-v}{1-\Omega}  \,, \quad & \text{if} \;\; \Omega <v\, \\
	\frac{1+v}{1+\Omega}  \,, \quad & \text{if} \;\; \Omega  \geq v \,\;\;
	\end{cases}\,
\end{eqnarray}
which is the same as the case in Einstein's gravity \cite{Mezei:2019dfv} and is always smaller than or equal to $\frac{2\pi}{\beta}$.

\item For $\mu <1 $, the OTOC depends on the relative values between $\Omega$ and $\mu$.
If $ \Omega <\mu$, we have
\begin{eqnarray}
	g(t,\varphi) = 1- \varepsilon_{VW} \begin{cases}
	-(1-\mu)\;
	e^{\lambda_+\left(t-\varphi\right)}   \,, \quad & \text{if} \;\; \Omega t <\varphi\, \\
	-2 \; e^{\lambda_m\big(t-\frac{\varphi}{v_m}\big)},\quad
	 & \text{if}\;\;  \Omega t  \geq \varphi \\
	\end{cases}\,.
	\end{eqnarray}
We have Lyapunov exponent $\lambda_m$ which satisfies the chaos bound while we have butterfly velocity $v_m<-1$ which exceeds the speed of light. The speed depend Lyapunov exponent \cite{Mezei:2019dfv} is 
\begin{eqnarray}
\label{eq:vdle2}
\lambda(v) = \frac{2\pi}{\beta} \begin{cases}
	\frac{1-v}{1-\Omega}  \,, \quad & \text{if} \;\; \Omega <v\, \\
	\frac{1-\mu\Omega+v(\mu-\Omega)}{1-\Omega^2}  \,, \quad & \text{if} \;\; \Omega  \geq v \,\;\;
	\end{cases}\,.
\end{eqnarray}
	
If $ \mu<\Omega$, the OTOC is given by
\begin{eqnarray}
	g(t,\varphi) = 1 -\varepsilon_{VW} \begin{cases}
	2\; 
	e^{\lambda_m\big(t-\frac{\varphi}{v_m}\big)}  \,, \quad & \text{if} \;\; \Omega t <\varphi\, \\
	(1+\mu)\;e^{\lambda_-\left(t+\varphi\right)}
	,\quad & \text{if}\;\; \Omega t  \geq \varphi \;\;
	\end{cases}\,.
\end{eqnarray}
Now the OTOC is controlled by Lyapunov exponents $\lambda_-$ and the butterfly velocity $v_-=-1$.  The speed depend Lyapunov exponent \cite{Mezei:2019dfv} is 
\begin{eqnarray}
\label{eq:vdle3}
\lambda(v) = \frac{2\pi}{\beta} \begin{cases}
	\frac{1-\mu\Omega+v(\mu-\Omega)}{1-\Omega^2}  \,, \quad & \text{if} \;\; \Omega <v\, \\
	\frac{1+v}{1+\Omega}  \,, \quad & \text{if} \;\; \Omega  \geq v \,\;\;
	\end{cases}\,
\end{eqnarray}	
In both cases (\ref{eq:vdle2}) and (\ref{eq:vdle3}), the speed depend Lyapunov exponents are  always smaller than or equal to $\frac{2\pi}{\beta}$.
	
\end{itemize}

In the above formulae, the correspondence of  the butterfly velocities (\ref{eq:vel-tmg}) and each Lyapunov exponents (\ref{eq:lya-tmg}) is verified. 

\subsubsection{OTOC at the chiral point $\mu=1$}
\label{sec:cp}

At the chiral point $\mu=1$ in TMG \cite{Li:2008dq}, due to the emergence of the log graviton, the dual field theory is proposed to be a rank-2 log CFT under certain boundary condition \cite{Grumiller:2008qz, Skenderis:2009nt}. For the calculations of OTOC, 
the case of $\mu = 1$ has to be treated separately since two solutions in (\ref{eq:swsolution}) for shock wave equation (\ref{eq:swequation}) degenerate and a new solution emerges. In the planar black hole with noncompact $\phi$, the profile of the shock wave is given by (\ref{eq:planarsw4}). If $\phi$ is periodic, for $\phi\in [0, 2\pi)$ we have 
\begin{eqnarray}
\begin{split}
h(\phi)&=
 \frac{\#}{4r_+}\left[\sum_{n=-\infty}^{0}  \left((1+2r_+ (\phi-2\pi n))\;e^{ \frac{-2\pi(\phi-2\pi n)}{\beta(1+\Omega)}} \right)
+\frac{1}{r_+}\; \sum_{n=1}^\infty e^{\frac{2\pi(\phi-2\pi n)}{\beta(1-\Omega)}}\right]   \\
&= \frac{\#}{4r_+}\left[\left( \frac{1+2 r_+\phi}{1-e^{-\frac{4\pi^2}{\beta (1+\Omega)}}}+\frac{4\pi r_+ e^\frac{4\pi^2}{\beta(1+\Omega)}}{(e^\frac{4\pi^2}{\beta(1+\Omega)}-1)^2}
\right)e^{ \frac{-2\pi\phi}{\beta(1+\Omega)}} 
+\frac{1}{r_+}\frac{1}{e^{\frac{4\pi^2}{\beta (1-\Omega)}}-1}e^{\frac{2\pi\phi}{\beta(1-\Omega)}} 
\right]\,.
\end{split}
\end{eqnarray}

Note that the term of form $\phi e^{ \frac{-2\pi\phi}{\beta(1+\Omega)}} $ is not the conventional term appearing in OTOC. Naively we have the Lyapunove exponents $\lambda_\pm$ with butterfly velocities $v_\pm$ as shown in (\ref{eq:lya-tmg}) and (\ref{eq:vel-tmg}). 
The calculation of the instantaneous Lyapunov exponent and OTOC proceed as before. In the regime $t\gg\beta$ and in the high temperature limit the instantaneous Lyapunov exponents takes the form 
\begin{eqnarray}
\lambda_\text{inst.}=\begin{cases}
\lambda_-  \,, \quad &\text{if} \;\;  t \in\big[0, ~ \frac{\pi (1+\Omega)}{\Omega}\big)\\
  \lambda_+\,, \quad &\text{if} \;\;  t \in\big[ \frac{\pi (1+\Omega)}{\Omega}, ~
  \frac{2\pi}{\Omega}\big)\\
	\end{cases}\,~~
\end{eqnarray}
and for the average Lyapunov exponents, we have $\langle\lambda_{\rm\;inst.}\rangle =\frac{2\pi}{\beta}$. 
The OTOC takes the following form in the high temperature limit and $|\Omega t-\varphi |\ll 1$
\begin{eqnarray}
	g(t,\varphi) = 1 -\varepsilon_{VW} \begin{cases}
	\frac{\#}{4r_+^2}\;
	e^{\lambda_+\left(t-\varphi\right)}  \,, \quad & \text{if} \;\; \Omega t <\varphi\, \\
	\frac{\#}{4r_+}\; e^{\lambda_-(t+\varphi)} \,, \quad & \text{if} \;\; \Omega t \geq \varphi \,\;\;
	\end{cases}\,.
	\end{eqnarray}
Thus at the chiral point the Lyapunov exponent is given by $\lambda_-$, satisfying the chaos bound, and the butterfly velocity is equal to the speed of light $v_-=-1.$
\vskip 0.4cm
From the discussions in subsections \ref{sec:lyexotoc} and \ref{sec:cp}, 
one can see the for any value of $\mu$, the Lyapunov exponents for systems dual to rotating BTZ in TMG are less than $\frac{2\pi}{\beta}$ and therefore the dual system is always not maximally chaotic. The average value of the instantaneous Lyapunov exponents always saturates the MSS chaos bound.  These properties are quite similar to Einstein gravity \cite{Mezei:2019dfv}. 
When $\mu<1$, 
the butterfly velocity can be faster than the speed of light. While when $\mu \geq 1$ the butterfly velocity is always equal to the speed of light.  Thus from the perspectives of speed bound of butterfly velocity, the physical allowed regime for the Chern-Simons coupling in TMG is $\mu\geq 1$. In this case, if we take the high temperature limit of the two independent instantaneous Lyapunov exponents as we analyzed in subsection \ref{sec:instan}, we found that they behave as step functions and take values among $\lambda_\pm$ 
as defined in (\ref{eq:lya-tmg}). 
From the analysis in subsection \ref{sec:lyexotoc}, it is clear that in the physical regime $\mu \geq 1$, the Lyapunov exponents and their corresponding butterfly velocities (\ref{eq:vel-tmg}) ``partly" matches the pole-skipping expectations as we show in the next parts of this section.  

\subsection{Pole-skipping from near horizon}
As emphasized in the beginning, pole-skipping is yet another signature of quantum chaos in many-body systems. Therefore it befits us to explore this phenomenon in systems that are not described by Einstein's gravity in order to better understand its connection to the chaos parameters obtained via OTOC. In the subsequent subsections, we study the pole-skipping phenomenon for rotating BTZ in TMG using three different ways to discuss the relations between these points and the Lyapunov exponents and butterfly velocities obtained from subsection \ref{sec-otoc}. In this and the following two subsections we treat the angular coordinate noncompact, i.e. we work in the high temperature limit. 

We proceed in a similar fashion as Einstein gravity, where we first study the pole-skipping from the near horizon equation of motion. We shall consider the linearized fluctuations of the metric around the black hole background. As before, equations of motion for the fluctuations are obtained by expanding the metric as $g_{ab} = {\bar g}_{ab} +h_{ab}$, and are given by
\begin{eqnarray}\label{tmgfluctuationeom}
R_{ab}^L-\frac{1}{2}{\bar g}_{ab}R^L + h_{ab}+\frac{1}{\mu}C_{ab}^L=0
\end{eqnarray}
where the linearized tensors are expressed as
\begin{eqnarray}
\begin{split}
R^L_{ab}&= \nabla^{c} \nabla_{(a}h_{b)c} - \frac{1}{2} \nabla^{c}\nabla_{c} h_{ab} - \frac{1}{2} \nabla_{a} \nabla_{b} h\,,\\
R^L&=\nabla^{a} \nabla^{b}h_{ab} -\nabla^{2} h+2h\,,\\
C_{ab}^L&=\varepsilon_a^{~cd}\nabla_c S_{db}^L+\frac{1}{2}\varepsilon_a^{~cd}\nabla_c h_{db}\,,\\
S_{ab}^L &= R^L_{ab} - \frac{1}{4} R^L \bar{g}_{ab}\,.
\end{split}
\end{eqnarray}
Once again we note that the above equations can be obtained by expanding the field equations \eqref{tmgeom} to linear order in $h_{ab}$ or to quadratic order in the action \eqref{tmgfullaction} and then varying it w.r.t. the fields.

We consider the $vv$-component of the equations of motion in the ingoing Eddington-Finkelstein coordinates of \eqref{eq:ingoingEF} and study its ``breakdown" near the horizon at special values of $\omega$ and $k$. This computation is completely analogous to its Einstein gravity counterpart and in this case the near horizon equation is of the form
\begin{align}\label{tmgevveom}
\begin{split}
 e^{(0)}_{vv} h^{(0)}_{vv} + e^{(0)}_{vr}h^{(0)}_{vr}+e^{(0)}_{v\phi}h^{(0)}_{v\phi}+e^{(0)}_{r\phi}h^{(0)}_{r\phi} + e^{(0)}_{\phi \phi}h^{(0)}_{\phi\phi} + e^{(1)}_{vv}h^{(1)}_{vv} + e^{(1)}_{v\phi}h^{(1)}_{v\phi}=0\,.
\end{split}
\end{align}
The coefficients $e^{(i)}_{ab}$ are functions of $k$, $\omega$, $r_+$ and $r_-$ and exact expressions for these coefficients can be found in appendix \ref{Evvcoeffs}. It can be checked that \eqref{tmgevveom} degenerates, \textit{i.e.} all the coefficients $e^{(i)}_{ab}$ vanishes at
\begin{eqnarray}\label{eq:pole-kerr-tmg}
(\omega, k) = \left(\frac{2\pi i}{\beta}\,, ~\pm \frac{2\pi}{i\beta (1 \mp \Omega)}\right)~~ \&~~~\left(\frac{2\pi i}{\beta}\,, ~\frac{2i\pi  (\Omega-\mu)}{\beta (1-\Omega^2)}\right)\,.
\end{eqnarray}

These points in the Schwarzschild coordinates (\ref{eq:metricbtz}) are located at 
\be\label{eq:pstmg}
\left(\frac{2\pi i}{\beta (1\mp \Omega)}\,,~~\mp\frac{2\pi i}{\beta (1\mp\Omega)}\right)~~\&~~
\left(\frac{2\pi i (1-\Omega\mu)}{\beta (1- \Omega^2)}\,,~~\frac{2i\pi  (\Omega-\mu)}{\beta (1-\Omega^2)}\right)\,.
\ee
The first set of points is familiar from Einstein gravity, corresponding to the two massless modes while the second point is the new pole-skipping point which corresponds to the massive graviton that arise in TMG.
From these pole-skipping points, we see that Lyapunov exponents are precisely of form (\ref{eq:lya-tmg}) and the butterfly velocities are (\ref{eq:vel-tmg}). For the case $\mu\geq 1$, these three Lyapunov exponents and butterfly velocities from pole-skipping take the values of step functions of instantaneous Lyapunov exponents from OTOC in the high temperature limit.

\subsection{Pole-skipping from holographic massive mode}
\label{sec:pshmm}
In this subsection, we calculate the additional  
pole-skipping point comparing Einstein gravity through a direct (numerical) computation of the retarded Green's function for the massive graviton mode from holography. 
The massive graviton mode which appears in the TMG is interpreted as the appearance of a new spin-2 operator in the dual CFT, which is denoted $t_{ij}$ and the pole-skipping point related to 
$(\lambda_m, v_m)$ in (\ref{eq:lya-tmg}) and (\ref{eq:vel-tmg}) 
is associated with the retarded correlator of this new operator. 

We work in the coordinates  (\ref{poincarebhmetric}). Note that the temperature of BTZ has been set to be $1/\pi$ in these coordinates and after the calculations we will transform back to the coordinates in metric (\ref{eq:metricbtz}) to compare with the results obtained in other subsections. 
 We can ascertain the form of the two-point function from the near boundary solution of the equations of motion in the radial gauge $h_{\rho a} = 0$. Near the boundary we have an expansion of the form \cite{Skenderis:2009nt},
\begin{eqnarray}
\label{nearbdyseries}
\begin{split}
h_{ij}(\rho) &= e^{-i\omega T + i k X}\Big[ h^{(0)}_{ij} + \rho  h^{(1)}_{ij} + \rho^2 h^{(2)}_{ij} + \rho^{-\delta}\big(b^{(0)}_{ij}  + \rho b^{(1)}_{ij} + \rho^2 b^{(2)}_{ij} + \cdots\big)\\ &\;\;\;\;\;\;\;\;\;\;\;\; + \rho^{\delta + 1}\big(c^{(0)}_{ij}  + \rho c^{(1)}_{ij} + \rho^2 c^{(2)}_{ij} + \cdots\big) \Big]
\end{split}
\end{eqnarray}
where we have taken the background metric to be \eqref{poincarebhmetric} and TMG coupling to be 
\begin{eqnarray}
\label{defdelta}
\mu = 2\delta+1\,.
\end{eqnarray}
It is easy to see that $h^{(0)}_{ij}$ acts as sources for the usual massless gravitons while $b^{(0)}_{ij}$ are the sources for the new massive gravitons. The coefficients $c^{(0)}_{ij}$ are the response terms associated with $b^{(0)}_{ij}$. This fact can be verified through an explicit computation of on-shell action following the holographic renormalization program of \cite{Skenderis:2009nt}.

It should be noted that the $h^{(a)}_{ij}$ coefficients are the solutions of Einstein gravity  and also satisfy TMG equations of motion. Since these are pure gauge modes, we cannot define a ingoing/outgoing waves for these modes at the horizon. The discussions of these massless modes should be the same as the ones in Einstein gravity in subsection \ref{sec:e2pchol} and in this subsection we shall focus only on the massive graviton mode. 

The $TT$, $TX$ and $XX$ components of the massive graviton mode are not independent and are constrained by the equations of motion, where leading $b$ and $c$ coefficients satisfy
\begin{align}
\begin{split}
b^{(0)}_{TT} = b^{(0)}_{TX} = b^{(0)}_{XX} \,,~~~~
c^{(0)}_{TT} = -c^{(0)}_{TX} = c^{(0)}_{XX}\,.
\end{split}
\end{align}
This shows that there is only one massive mode and the correlator is simply given by
$G^{R} \propto \frac{c^{(0)}}{b^{(0)}}$ 
where we have ignored possible normalization constants. Pole-skipping can be shown by finding the common 
zeros of $b^{(0)}_{TT}$ and $c^{(0)}_{TT}$ for purely imaginary values of $\omega$ and $k$. In Lorentzian AdS
the retarded correlator $\langle t_{00}(t,x)t_{00}(0,0)\rangle_R$ is obtained by imposing ingoing wave condition at the horizon \cite{Son:2002sd}. The ingoing wave solution has a near horizon expansion which is given by
\begin{align}\label{nearhorizonexp}
\begin{split}
h_{TT}(\rho) &= (1-\rho)^{4-\frac{i\omega}{2}}\left( \frac{4+k^2 - 4i k \mu -4\mu^2 +2i\omega}{16(\omega+4i)(\omega+6i)} + \mathcal{O}(1-\rho) \right)\,, \\
h_{TX}(\rho) &= (1-\rho)^{2-\frac{i\omega}{2}}\left( \frac{k-2i\mu}{4(\omega+4i)} + \mathcal{O}(1-\rho) \right)\,, \\
h_{XX}(\rho) &=  (1-\rho)^{-\frac{i\omega}{2}}\left( 1 - \left(1+\frac{i\omega}{4}\right)(1-\rho) + \mathcal{O}((1-\rho)^2) \right)\,.
\end{split}
\end{align} 
We see that once we choose the normalization to be $h^{(0)}_{XX} = 1$, there are no undetermined coefficients in the expansion \eqref{nearhorizonexp}. 
We use the shooting method to generate the solutions of the TMG equations of motion with above boundary conditions at the horizon. The coefficients $b^{(0)}_{ij}$ and $c^{(0)}_{ij}$ can be extracted from the solutions $h_{ij}$ by inverting the series \eqref{nearbdyseries} upto desired accuracy.

For pole-skipping, we have to analyze the zeros of $b^{(0)}_{TT}$ and $c^{(0)}_{TT}$ coefficients for purely imaginary values of $\omega$ and  $k$. The zeros of $b^0_{tt}$ becomes the poles of the Green's function  
while the zeros of $c^{(0)}_{tt}$ becomes the zeros of the Green's function. Our numerical scan reveals a line of zeros for $b^{(0)}_{tt}$ coefficient which fit the analytic formula
\begin{eqnarray}\label{btt0line}
\text{Im} \omega =\text{Im} k -4\delta
\end{eqnarray}
which is consistent with the  QNMs found in \cite{Sachs:2008gt} when analytically continued to imaginary $\omega$ and $k$. The zeros of $c^{(0)}_{tt}$ also falls on a line, which is fitted by the line
\begin{eqnarray}\label{ctt0line}
\text{Im} \omega = -\text{Im} k + 4(1+\delta)\,.
\end{eqnarray}

\begin{figure}[h]
\centering
\includegraphics[width=0.6\textwidth]{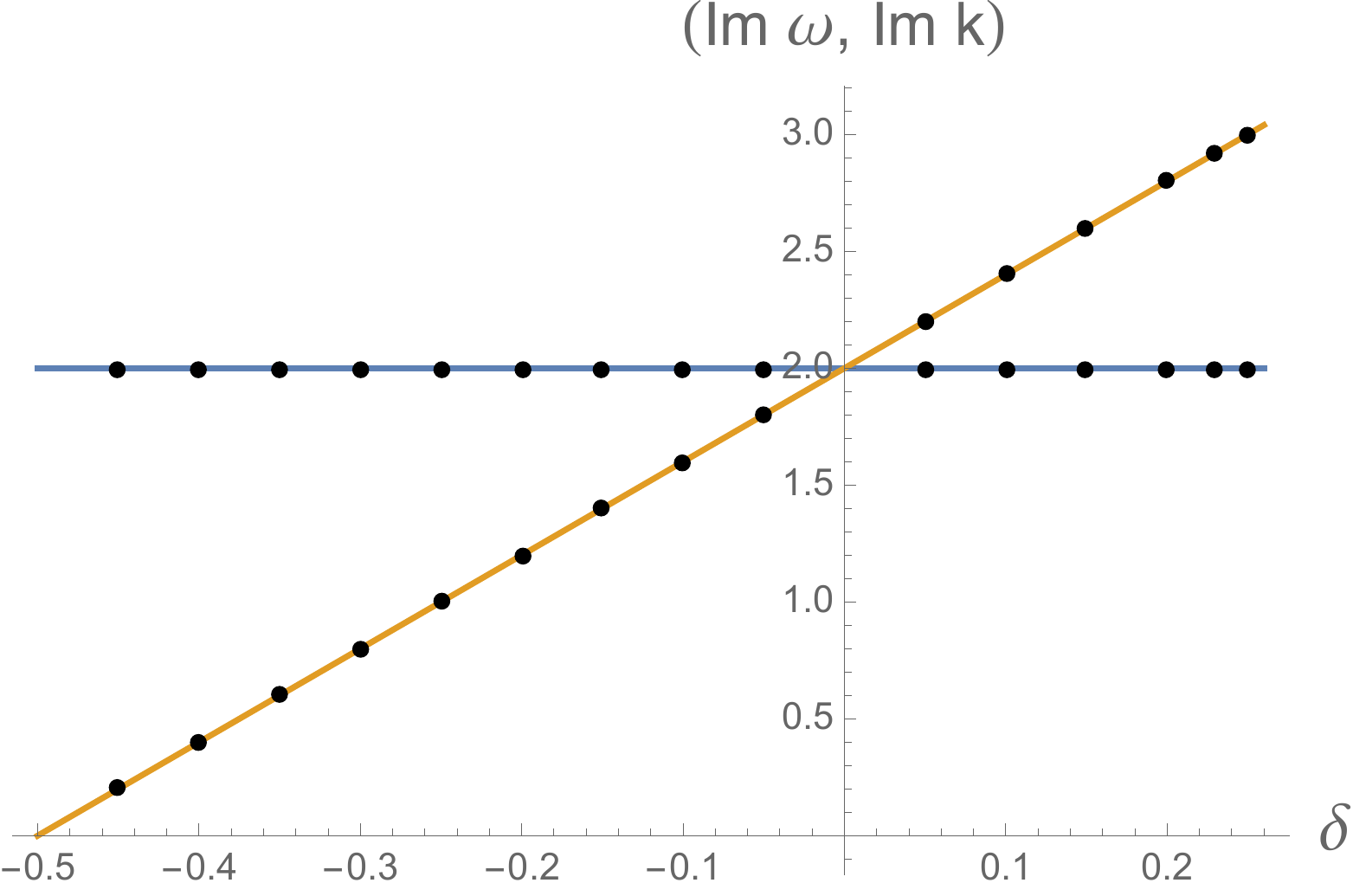}
\caption{\small The plot shows the numerical fitting for pole-skipping $\text{Im} \omega$ (blue) and $\text{Im} k$ (orange) as a function of $\delta$. 
The solid lines corresponds to the analytic curve $(\text{Im} \omega, \text{Im} k) = (2,2(1+2\delta))$ while the dots are our numerical results for various $\delta$.}
\label{fig:pspoints}
 \end{figure}

We have limited our analysis to the above two lines because these are enough to reproduce the near-horizon results. However the scanned space of pure imaginary $(\omega,k)$ also shows additional lines of zeros for both $b^{(0)}_{tt}$ and $c^{(0)}_{tt}$ which we will not discuss in this work. These lines may be important to explain additional pole-skipping points that appear in the field theory analysis. For the lines \eqref{btt0line} and \eqref{ctt0line}, the pole-skipping point is given by their intersection\footnote{Note that this is the result in the coordinates \eqref{poincarebhmetric} where the temperature is $1/\pi$. One can easily transform the results here back to the Schwarzschild coordinate following the discussions in the appendix \ref{app:btzPoincare}. It would be interesting to analytically obtain the pole-skipping point.}
\begin{eqnarray}
\label{eq:skippedpo}
\omega = 2i\,, \quad k = 2i(1+2\delta)
\end{eqnarray}
as show in figure~\ref{fig:pspoints}.

Using the transformation of momentum space variables \eqref{momentum-transformation} and (\ref{defdelta}), we can easily verify that this point corresponds precisely to the parameters related to massive mode in \eqref{eq:pstmg} coming from the near-horizon analysis of the equations of motion.

\subsection{Pole-skipping from CFT analysis}
One can also obtain compelling hints for pole-skipping from generic CFT methods. In this subsection, we extend the calculation of CFT two-point function to also include the dual operator for the massive mode. From the analysis of \cite{Skenderis:2009nt}, it was concluded that the CFT contains operators $T_{zz}$, $T_{\bar{z}\bar{z}}$ and $X_{zz}$ which are dual to the massless and massive graviton modes respectively, and their corresponding (Euclidean) two-point functions were also calculated. The operators  $T_{zz}$ and $T_{\bar{z}\bar{z}}$ corresponds to the usual stress tensor of the CFT and are of holomorphic dimensions $(2,0)$ and $ (0,2)$. However, unlike CFT dual to Einstein gravity, in TMG the left and right CFTs carry different central charge as shown in (\ref{eq:TMGcc}).

The operator $X_{zz}$, dual to the massive mode is non-chiral with dimensions $(2+\delta,\delta)$ and its two-point function on the plane is
\begin{eqnarray}
\langle X_{zz}(z,\bar{z}) X_{zz}(0,0)\rangle = \frac{\mathcal{C}_\delta}{z^{2(2+\delta)}\bar{z}^{2\delta}}
\end{eqnarray}
where $\delta$ is as defined in \eqref{defdelta} and $\mathcal{C}_\delta$  is a normalization constant which is irrelevant for our purposes. To study pole-skipping, we have to transform this Euclidean result to a Lorentzian correlator on cylinder. The calculation proceeds exactly as in the case of stress-tensor in subsection \ref{sec:rgfCFT} and we will  only list the main result here. 
The Fourier transform of $G(t,\sigma)$ is 
\begin{align}
\begin{split}
G(\omega,k) &\propto  \sinh\left[\frac{\beta_L}{2}\left(\frac{\omega+k}{2}\right)+ \frac{\beta_R}{2}\left(\frac{\omega-k}{2}\right) \right] \left|\Gamma\left(2+\delta+\frac{i \beta_L}{2\pi}\left(\frac{\omega+k}{2}\right)\right) \right|^2\times  \\ &~~~~~\times\left|\Gamma\left(\delta+\frac{i \beta_R}{2\pi}\left(\frac{\omega-k}{2}\right)\right) \right|^2\,.
\end{split}
\end{align}
Thus the retarded Green's function will now be, 
\be \label{eq:retardedG}
G_R(\omega, k)\sim i\pi G(\omega, k)+\int_{-\infty}^\infty d\lambda  \frac{G(\lambda, k)}{\lambda-\omega}\,.
\ee
If we restrict $\omega, k$ to be real, the first term is the imaginary part of the retarded Green's function and the second term is the real part which connects to the imaginary part by Kramers–Kronig relation. Similar structure for the retarded Green's function of scalar operator in CFT has been obtained in holography before, {\it e.g.} equation (4.16) in \cite{Son:2002sd}. The last term will modify the retarded Green's function. The integration in \eqref{eq:retardedG} is quite nontrivial, following \cite{Son:2002sd} it is natural to conjecture that it takes the form\footnote{Notice that when $\delta$ is an integer there are subtleties here and we do not consider them here.}
\begin{align}
\begin{split} 
G_R(\omega,k) &\propto
\sin\left[\delta+\frac{i \beta_R}{2\pi}\left(\frac{\omega-k}{2}\right) \right] \sin\left[2+\delta+\frac{i \beta_L}{2\pi}\left(\frac{\omega+k}{2}\right) \right]\times  \\ &~~~~~\times\left|\Gamma\left(\delta+\frac{i \beta_R}{2\pi}\left(\frac{\omega-k}{2}\right)\right) \right|^2 \left|\Gamma\left(2+\delta+\frac{i \beta_L}{2\pi}\left(\frac{\omega+k}{2}\right)\right) \right|^2\,.
\end{split}
\end{align}
Notice that when $\beta_L=\beta_R$, the above formula reduces to equation (4.16) in \cite{Son:2002sd}.

A convenient way to study the pole-skipping is in the boundary of Poincare-like coordinate in appendix \ref{app:btzPoincare}. We can transform back to Schwarzschild coordinate at the end of calculation. 
We need to analyze the zeros and poles of 
\begin{eqnarray}\label{gpgm}
\begin{split} 
G_R(\omega,k) &\propto \sin\big[\delta + \frac{i}{4}(\omega-k)\big] \left| \Gamma\left(\delta + \frac{i}{4}(\omega-k) \right) \right|^2\times\\
&~~~\times
\sin\big[ 2+\delta + \frac{i}{4}(\omega+k)\big] \left| \Gamma\left( 2+\delta + \frac{i}{4}(\omega+k) \right) \right|^2\,.
\end{split}
\end{eqnarray}
Notice that the poles of $\Gamma\left(\delta + \frac{i}{4}(\omega-k) \right) $ and $\Gamma\left( 2+\delta + \frac{i}{4}(\omega+k) \right)$ are not poles due to the term of 
$\sin\big[\delta + \frac{i}{4}(\omega-k)\big] \Gamma\left(\delta + \frac{i}{4}(\omega-k) \right)$ and $\sin\big[ 2+\delta + \frac{i}{4}(\omega+k)\big]  \Gamma\left( 2+\delta + \frac{i}{4}(\omega+k) \right)$. One can get the pole-skipping points are:  
$(\omega, k)=2i (2-n-m, 2+2\delta+n-m)$ or $(\omega, k)=2i (-2-n-m, -2-2\delta-n+m)$ where $n=0,1,2,\cdots$ and $m=1,2,\cdots$. The only pole appearing in the upper half plane is $(\omega, k)=(2i, 2i(1+2\delta))$\footnote{If $\mu\Omega$ is large enough there might be subtleties and we will not consider them here.}, {\it i.e.}  (\ref{eq:skippedpo}) in subsection \ref{sec:pshmm}. Using (\ref{momentum-transformation}) to transform back to Schwarzschild coordinate, we get the chaos parameters in \eqref{eq:pstmg}.

\section{Conclusion and Discussions}
\label{conclusion}

We discussed the relations between the quantum chaos parameters including Lyapunov exponents and butterfly velocities from OTOC and from pole-skipping in systems dual to rotating BTZ black holes in TMG, which turns out to be a non-maximally chaotic system at high temperature.\footnote{A recent study on pole-skipping in the retarded two-point correlator of energy density operators in non-maximally chaotic system can be found in \cite{Choi:2020tdj}. However, from our study, the pole-skipping in the retarded correlation function of the spin two operators which is dual to massive graviton also plays an important role for chaos.} From OTOC, we studied the behavior of  the 
instantaneous Lyapunov exponent. We also studied the chaos parameters using pole-skipping methods from the near horizon equation, from holographic two point correlators of energy densities, and from CFT techniques and found that these approaches gave consistent results.

For any value of the gravitational Chern-Simons coupling $\mu$, at high temperature 
the instantaneous Lyapunov exponents obtained 
from OTOC can only produce part of exponents from pole-skipping. 
When $\mu\geq 1$ the MSS chaos bound is satisfied and the butterfly velocity is equal to the speed of light, whereas in regime $\mu<1$ the Lyapunov exponent satisfies the  chaos bound while the butterfly velocity can be faster than the speed of light. Therefore the velocity bounds the Chern-Simons coupling to the regime $\mu\geq 1$. Intriguingly, in this regime the quantum chaos is well-behaved and the BTZ black holes are stable \cite{Park:2006gt, Li:2008dq}. In our calculation of pole-skipping we have assumed that the angular coordinate is noncompact which is a reasonable approximation at high temperature. It would be interesting to include the periodicity in pole-skipping calculations, {\it e.g.} to compute the retarded Green's function on a dual torus to compare the OTOC calculation.\footnote{Pole-skipping for CFT's with $c_L=c_R$ and $\beta_L=\beta_R$ on a torus with has been studied in \cite{Ramirez:2020qer} and it would be interesting to generalize to the cases with $c_L\neq c_R$ and $\beta_L\neq \beta_R$.}

There are some immediate generalizations of our work. For example, one can study the quantum chaos in higher dimensional rotating black holes using OTOC and pole-skipping to discuss the relations of chaotic parameters. One can also study systems with gravitational anomalies to see if there is any constraint on the couplings related to gravitational anomaly from the MSS chaos bound.  

\vspace{0.95cm}
\appendix
\addtocontents{toc}{\protect\setcounter{tocdepth}{0}}

\section{BTZ in Different Coordinates}\label{app:threecoords}
It is useful to express BTZ metric (\ref{eq:metricbtz}) or (\ref{eq:me-comoving}) in different coordinates for different purposes. In this appendix, we list the three useful coordinates which are put to use in this paper. It is convenient to calculate the holographic correlators in Poincare-like coordinate. We shall use the ingoing Eddington-Finkelstein coordinates to compute the pole-skipping from near horizon equation of motion. For the study of shock wave solution we shall use the Kruskal coordinate. After the calculations done in these coordinates, we will transform back to the metric (\ref{eq:metricbtz}) and study the chaos in  Schwarzschild coordinates. 

\subsection{Poincare-like Coordinates}
\label{app:btzPoincare}
The first coordinate system we introduce is a Poincare-like coordinates in which the BTZ metric has a Fefferman-Graham form. This coordinates lends itself to easy computation of holographic correlators and the results can be readily compared with \cite{Skenderis:2009nt}.
The metric (\ref{eq:metricbtz}) can be brought to a Poincare-like form, 
by introducing a new set of coordinates $(z,T,X)$ which is related to the Schwarzschild coordinates by,
\begin{align}
\begin{split}
\label{eq:coortrans}
r^2 &= r_+^2 + \left(\frac{r_+^2 - r_-^2}{4}\right)\left(\frac{1}{z}-z\right)^2\,, \\
t &= \left(\frac{2r_+}{r_+^2 - r_-^2}\right)T - \left(\frac{2r_-}{r_+^2 - r_-^2}\right)X\,, \\
\varphi &= \left(\frac{2r_-}{r_+^2 - r_-^2}\right)T - \left(\frac{2r_+}{r_+^2 - r_-^2}\right)X\,.
\end{split}
\end{align}
In this coordinates, the BTZ metric takes the form of a non-rotating black hole
\begin{eqnarray}
ds^2 = \frac{dz^2}{z^2}  + \frac{1}{z^2} \Big[-(1-z^2)^2 dT^2 + (1+z^2)^2 dX^2 \Big]
\end{eqnarray}
where $z=0$ is the boundary, $z=1$ is the location of the horizon\footnote{We could have left the Schwarzschild time and angular coordinates intact. However this makes calculations unnecessarily complicated and moreover one can always relate the Schwartzshild coordinates $(t,\varphi)$ and Poincare $(T,X)$ results by a linear transformation.} and the coordinates cover the region outside the outer horizon. Similar coordinates have been used before {\it e.g.} in \cite{Carlip:1995qv,Son:2002sd}. 

Since only even powers of $z$ appears in the above metric, we can define a new radial coordinate $\rho = z^2$ in which the metric is now of the form
\begin{eqnarray}\label{poincarebhmetric}
ds^2 = \frac{d\rho^2}{4\rho^2}  + \frac{1}{\rho} \Big[-(1-\rho)^2 dT^2 + (1+\rho)^2 dX^2 \Big]\,.
\end{eqnarray}
The temperature for this geometry is $1/\pi$.

This is the coordinate system we use to numerically analyse the retarded correlator for  the linearized fluctuations in TMG. It is also instructive to relate the momentum space variables between the Schwarzschild and Poincare coordinates, which transform as inverse of the transformation (\ref{eq:coortrans}) 
\begin{align}\label{momentum-transformation}
\begin{split}
\omega_{\rm sch} = \frac{r_+}{2}\omega - \frac{r_-}{2}k\,,~~~
k_{\rm sch} = \frac{r_-}{2}\omega - \frac{r_+}{2}k\,,
\end{split}
\end{align}
where $(\omega, k)$ are quantities in Poincare coordinates while $(\omega_{\rm sch}, k_{\rm sch})$ in Schwarzschild coordinates. 

\subsection{(Ingoing) Eddington-Finkelstein Coordinates}
\label{app:btzEF}

The ingoing Eddington-Finkelstein form of the metric can be obtained from the comoving metric by defining a new time coordinate $v=t+r_*$, where $r_*$ is the tortoise coordinate for \eqref{eq:me-comoving}
\begin{eqnarray}\label{tortoise}
r_*=\frac{r_+}{2(r_+^2-r_-^2)}\log \frac{\sqrt{r^2-r_-^2}-\sqrt{r_+^2-r_-^2}}{\sqrt{r^2-r_-^2}+\sqrt{r_+^2-r_-^2}}\,.
\end{eqnarray}
The metric \eqref{eq:me-comoving} now becomes
\begin{align}\label{eq:ingoingEF}
\begin{split}
ds^2 &= -\frac{(r^2-r_+^2)(r_+^2-r_-^2)}{r_+^2} dv^2 + 2 \frac{r}{r_+}\sqrt{\frac{r_+^2 - r_-^2}{r^2 - r_-^2}} dv dr +  \frac{2 r_-(r^2 - r_+^2)}{r_+}dv d\phi \\ &\hspace{6cm} - \frac{2 r r_{-}}{\sqrt{(r^2 - r_-^2)(r_+^2 - r_-^2)}}dr d\phi + r^2 d\phi^2\,.
\end{split}
\end{align}
Here $(v, r, \phi)$ are ingoing Eddington-Finkelstein coordinates.

\subsection{Kruskal Coordinates}\label{app:btzKru}
From the co-moving coordinates (\ref{eq:me-comoving}) we can obtain the maximally extended Kruskal coordinates through 
\begin{eqnarray}
U = -e^{-\kappa(t-r_*)}\,, \quad V= e^{\kappa(t+r_*)}\,,\qquad \kappa = \frac{r_{+}^{2} - r_{-}^{2}}{r_+}
\end{eqnarray}
where $\kappa$ is the surface gravity and $r_*$ is the tortoise coordinate \eqref{tortoise}. The Kruskal coordinates $(U,V)$ can be extended beyond the domain of its definition and covers the whole BTZ spacetime upto $r=0$. The metric \eqref{eq:me-comoving} now takes following form,
\be\label{eq:btzKruskal}
ds^2=\frac{-4 dUdV-4 r_-(UdV-VdU)d\phi+\big[(1-UV)^2r_+^2+4UV r_-^2\big]d\phi^2}{(1+UV)^2}\,.
\ee

\section{Coefficients in (\ref{PoincareEinEucSoln})}
\label{app:coe3.13}
The coefficients appearing in (\ref{PoincareEinEucSoln}) are of the following form 
\begin{align}
\begin{split}
\tilde{h}^{(1)}_{\tau \tau} &= -\frac{1}{2(\omega^2_E+k^2)}\left[ \omega_E^2(4+k^2)\tilde{h}^{(0)}_{\tau\tau} + 2\omega_E(4-\omega_E^2)\tilde{h}^{(0)}_{\tau X} - \omega_E^2(4-\omega_E^2)\tilde{h}^{(0)}_{XX}\right] \\
\tilde{h}^{(1)}_{\tau X} &= -\frac{1}{2(\omega^2_E+k^2)}\left[ k\omega_E(4+k^2)\tilde{h}^{(0)}_{\tau\tau} - 2(2k^2 - 2\omega_E^2- k^2\omega_E^2)\tilde{h}^{(0)}_{\tau X} - k\omega_E(4-\omega_E^2)\tilde{h}^{(0)}_{XX}\right] \\
\tilde{h}^{(1)}_{XX} &= -\frac{1}{2(\omega^2_E+k^2)}\left[ k^2(4+k^2)\tilde{h}^{(0)}_{\tau\tau} - 2k\omega_E(4+ k^2)\tilde{h}^{(0)}_{\tau X} - k^2(4-\omega_E^2)\tilde{h}^{(0)}_{XX}\right] 
\end{split}
\end{align}
and 
\begin{align}
\begin{split}
\tilde{h}^{(2)}_{\tau \tau} &= -\frac{1}{2(\omega^2_E+k^2)}\left[ (2k^2-2\omega_E^2 - k^2\omega_E^2)\tilde{h}^{(0)}_{\tau\tau} - 2k\omega_E(4-\omega_E^2)\tilde{h}^{(0)}_{\tau X} + \omega_E^2(4-\omega_E^2)\tilde{h}^{(0)}_{XX}\right] \\
\tilde{h}^{(2)}_{\tau X} &= \tilde{h}^{(0)}_{\tau X} \\
\tilde{h}^{(1)}_{XX} &= -\frac{1}{2(\omega^2_E+k^2)}\left[ k^2(4+k^2)\tilde{h}^0_{\tau\tau} - 2k\omega_E(4+ k^2)\tilde{h}^{(0)}_{\tau X} - (2k^2 - 2\omega_E^2 - k^2\omega_E^2)\tilde{h}^{(0)}_{XX}\right] \,.
\end{split}
\end{align}

\section{Coefficients of Leading Order $E_{vv}$ in TMG}\label{Evvcoeffs}
The coefficients that appear in the leading order $E_{vv}$ equation \eqref{tmgevveom} are given by 
\begin{align}
\begin{split}
e^{(0)}_{vv} &= 2\pi\beta^2 \omega \left(i\beta \omega \Omega^3 + 2\pi \Omega - 2\pi \mu (1-\Omega^2) \right) - k^3 \beta^4(1-\Omega^2)^3  \\ &\qquad + 2\pi \beta^2 k (1-\Omega^2)\left(2\pi(1-2\mu \Omega) + i\beta \omega (2-3\Omega^2)\right) \\ &\qquad- k^2 \beta^3 (1-\Omega)^2 \left(2\pi i (\mu - 2 \Omega) - \beta\omega\Omega \right)\\
e^{(0)}_{vr} &= 4\pi^2 (2\pi + i\beta \omega)\left( 4\pi \omega \Omega - k(4\pi + i\beta \omega)(1-\Omega^2) \right) \\
e^{(0)}_{v\phi} &= 2\beta^2(2\pi+i\beta\omega)(1-\Omega^2)^2\left(2\pi\omega -k(2\pi\mu + i\beta \omega\Omega)+ i k^2 \beta(1-\Omega^2)  \right)\\
e^{(0)}_{r\phi} &= -4\pi^2 i \beta \omega^2 (2\pi+i \beta\omega)(1-\Omega^2)\\
e^{(0)}_{\phi\phi} &= \omega \beta^2(2\pi i -\beta\omega)(1-\Omega^2)^2 \left( 2\pi i \mu - \beta \omega \mu - k\beta(1-\Omega^2) \right)\\
e^{(1)}_{vv} &= -4k\pi^2 \beta (2\pi + i\beta \omega)(1-\Omega^2) \\
e^{(1)}_{v\phi} &= -4\omega\pi^2 \beta(2\pi + i \beta \omega)(1-\Omega^2)
\end{split}
\end{align}
Here we have traded the dependence on $r_+$, $r_-$ for $\beta$, $\Omega$ using the relations,
\begin{eqnarray}
r_+ = \frac{2\pi}{\beta(1-\Omega^2)}\,, \qquad r_- = \frac{2\pi\Omega}{\beta(1-\Omega^2)}\,.
\end{eqnarray}

\subsection*{Acknowledgments}
We are grateful to Rong-Gen Cai, Shu Lin, Gaurav Narain and Ya-Wen Sun  for useful discussions, to Viktor Jahnke, Keun-Young Kim for helpful correspondence, and the anonymous referee for helpful comments. This work is supported by the National Natural Science Foundation of China grant No.11875083. A.R is also supported by Zhuoyue Postdoc Fellowship of Beihang University (ZYBH2018-01).

\end{document}